\documentclass[letterpaper,journal]{IEEEtran}
\usepackage{amsmath,amssymb,amsfonts}
\usepackage{array}
\usepackage{textcomp}
\usepackage[hyphens]{url}
\usepackage{graphicx}
\usepackage{cite}
\newcommand{\panel}[3]{\begin{minipage}[c]{#1}\centering\includegraphics[width=\linewidth]{#2}\\[1pt]{\footnotesize (#3)}\end{minipage}}
\newcommand{\tabhead}[1]{{\renewcommand{\arraystretch}{1}\begin{tabular}[c]{@{}>{\bfseries\boldmath}c@{}}#1\end{tabular}}}

\begin{document}

\title{Emerging Vertical Power Delivery in AI Data Centers: Challenges and Solutions}

\author{Charles~Isaac~Zahner,~\IEEEmembership{Graduate Student Member,~IEEE,}
Amirhossein~Habibi,~\IEEEmembership{Graduate Student Member,~IEEE,}
and~Mehdi~Ferdowsi,~\IEEEmembership{Fellow,~IEEE}
\thanks{The authors are with the Department of Electrical and Computer Engineering, Missouri University of Science and Technology, Rolla, MO 65409 USA.}
\thanks}

\markboth{}{Zahner \MakeLowercase{\textit{et al.}}: Emerging Vertical Power Delivery in AI Data Centers: Challenges and Solutions}

\maketitle

\begin{abstract}
The rapid growth of artificial intelligence (AI) workloads is increasing power demand in AI data centers, and processor current is projected to reach 2,000~A and beyond. This trend is exposing the limitations of lateral power delivery (LPD) to processors and has motivated growing interest in vertical power delivery (VPD). This paper reviews the power delivery network (PDN) of AI data centers and the upcoming transition from LPD to VPD. First, the 800~VDC power delivery architecture, the impedance behavior of the PDN, and the limitations of LPD are discussed. Then, VPD implementation approaches, reported converter designs, and packaging technologies are examined, together with integrated voltage regulators (IVRs) and embedded passive components. In addition, power integrity simulations are presented to compare regulator and plane positions, from which a set of PDN design guidelines is derived. The results show that moving a voltage regulator module (VRM) from 50~mm away to directly below the die reduces the loop inductance by nearly half. In a complete PDN, this relocation lowers the low-frequency anti-resonant peak, whereas the mid- and high-frequency impedance depends on the decoupling stages and the position of the power and ground planes. This paper also shows that achieving practical VPD remains closely tied to challenges such as regulator height, passive-component embedding, and thermal management.
\end{abstract}


\section{Introduction}
\label{sec:intro}
\IEEEPARstart{A}{rtificial} intelligence (AI) is regarded as one of the most influential technologies of the 21st century, with major economic and social effects expected as its use continues to grow \cite{b1,b2}. Between 2025 and 2030, the global AI data center market is forecast to expand at a compound annual growth rate (CAGR) of 28.3\%, with the largest share held by North America \cite{b3}. This expansion comes mainly from generative AI, whose rise was made possible by hardware progress and breakthroughs in model architectures and algorithms \cite{b2}. The result has been a rapid increase in AI workloads, and its electrical consequence is a sharp rise in data center power demand \cite{b2,b4,b5}. In the United States, data centers were estimated to consume approximately 180~TWh in 2024, which corresponds to 4.4\% of the national electricity consumption \cite{b1}. Nearly half of the growth in U.S. electricity demand up to 2030 is expected to come from the continued expansion of these facilities \cite{b1}.

\begin{figure*}[!t]
\centerline{\includegraphics[width=0.86\textwidth]{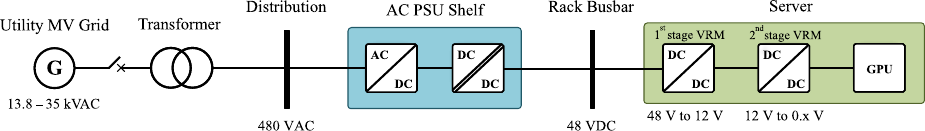}}
\caption{Current power delivery architecture with 48~VDC rack busbar distribution for AI data centers.\label{fig:1}}
\end{figure*}

AI data centers' power delivery has long been based on the architecture illustrated in Fig.~\ref{fig:1} \cite{b6}. The graphics processing units (GPUs) that execute AI workloads are the target load of this power delivery chain \cite{b7,b8,b9,b10,b11,b12}. They are fed from the grid through different power converter modules, including power supply units (PSUs) and voltage regulator modules (VRMs) \cite{b13,b14}. Each rack is fed from a 400/415/480~VAC busbar, and an in-rack PSU converts this input into an intermediate bus voltage (IBV) that typically lies in the 48--54~VDC range \cite{b15,b16}. The PSU itself is a device that converts an AC or a DC input into one or more DC outputs \cite{b3}. It is commonly built from two conversion stages: a first stage that performs power factor correction (PFC) and establishes a bus of approximately 400~VDC, and a second, isolated DC--DC stage that converts this bus into the unregulated IBV supplied to the servers \cite{b17}. The VRM is the last conversion stage of the power delivery chain and the one closest to the GPU \cite{b18}. Since the output voltage of the VRM stage is much lower than that of the previous stages, the required power has to be delivered at a much higher current, and the current demand placed on this stage is therefore severe. In the early 2020s, GPUs were already projected to begin drawing more than 1,000~A \cite{b19,b20}. Since then, the supply voltage has fallen from around 1~V to approximately 0.6--0.8~V, and also the technology node has shrunk from about 90~nm to about 5~nm \cite{b21,b22,b23}. As a result, processor current is projected to reach 2,000~A and beyond by 2028, with current densities of 3~A/mm$^{2}$ and above \cite{b24,b25,b26}. Hence, the design of the VRM is limited by two requirements. First, it must be physically small, and second, it must be able to follow fast load transients \cite{b20}. This combination of high static current and fast, large-amplitude transients has consequently made the reduction of power delivery network (PDN) losses a primary design objective in AI data centers \cite{b27,b28,b29}.

The VRM in the 480~VAC architecture is normally realized in two stages that convert 48~V to below 1~V through an intermediate 12~V rail, consisting of a high-ratio voltage converter followed by a buck converter \cite{b20}. In future AI data center implementations, and also in the newly introduced 800~VDC power delivery architecture, the two-stage VRM is going to be replaced by a single-stage VRM that converts 48~V directly to the core voltage \cite{b20}. In the systems deployed today, the VRM stage is placed beside the processor package in the server, and the low-voltage, high-current output is carried laterally across the board to feed the load \cite{b30}. This arrangement, referred to as lateral power delivery (LPD), is the most widely used method of supplying GPUs. Two limitations follow from this arrangement. First, since the current carried along the lateral path is high, even a small path resistance produces a routing loss that is unacceptably large \cite{b20}. Second, the large output filter inductors and capacitors, together with the regulator itself, occupy a substantial part of the board area, which has to be provided next to the processor package, where space is already scarce. As load currents continue to rise, both limitations become progressively more severe, and neither of them can be removed by improving the converters alone.

Overcoming these limitations requires the VRM to be brought closer to the GPU in order to shorten the PDN path and reduce its losses, which improves overall system efficiency \cite{b31,b32}. Vertical power delivery (VPD) is a promising solution to these challenges, in which the VRMs are mounted vertically below the load so that the PDN path is shortened. In another implementation of VPD, it is more efficient to mount a number of VRMs on the motherboard vertically below the processor, and to have them supply integrated voltage regulators (IVRs) positioned even closer to the die \cite{b20}. IVRs have long been used to address these challenges in high-performance computing (HPC) applications, and they have now become a subject of interest for incorporation into VPD in AI data center applications \cite{b33,b34}. Implementing these approaches requires VRMs and the PDN to be re-examined jointly, so that the limited space available around the processor and the loss along the delivery path are addressed together in the pursuit of higher efficiency \cite{b30}. Accordingly, this paper discusses the PDN issues associated with AI data centers and the limitations of LPD. In addition, the transition toward VPD is described, together with its implementation challenges and design considerations. IVR and embedded passive technologies that make vertical delivery practical are also presented. Finally, a set of PDN design rules and considerations for VPD is outlined, derived from PDN circuit design principles and their governing equations, and verified through simulation.

The organization of this paper is as follows. In Section~\ref{sec:800v}, the 800~VDC power delivery architecture is reviewed. In Section~\ref{sec:pdn}, the PDN and its components are described. In Section~\ref{sec:lpd}, LPD and its limitations are discussed, and in Section~\ref{sec:vpd}, VPD and the related configurations are presented. In Section~\ref{sec:ivr}, IVRs, fully integrated voltage regulators (FIVRs), and the passive-component solutions that enable component embedding are reviewed. In Section~\ref{sec:sim}, Cadence power integrity simulations are presented, and a discussion is provided to set PDN design rules and considerations. Finally, Section~\ref{sec:conc} concludes the paper.

\begin{figure*}[!t]
\centerline{\includegraphics[width=0.86\textwidth]{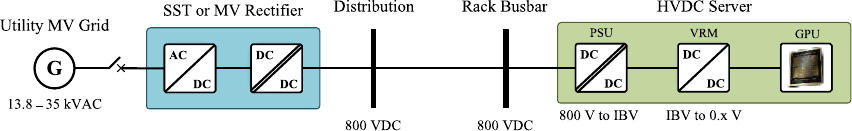}}
\caption{800~VDC power delivery architecture for next-generation AI data centers.\label{fig:2}}
\end{figure*}

\section{800 VDC Architecture}
\label{sec:800v}
The rapid growth of AI workloads is increasing rack power demand in AI data centers. Today's information technology (IT) racks typically operate at power levels in the 10--30~kW class \cite{b6}. Emerging rack-scale AI systems such as NVIDIA GB200 NVL72 and NVIDIA GB300 NVL72 \cite{b35,b36} are pushing rack power into the 120~kW class, and NVIDIA's roadmap toward Kyber racks powered by Rubin Ultra GPUs is targeting $>$1~MW class racks \cite{b37,b38}. This trend is exposing the limitations of the current 48~VDC rack distribution architecture and has motivated growing interest in higher-voltage power delivery architectures \cite{b13}. To address this challenge, the 800~VDC architecture was introduced by NVIDIA in 2025 for next-generation power distribution in AI data centers \cite{b6,b37}. A detailed review of the 800~VDC architecture and of the isolated 800~V to 48~V converters proposed for it is given in \cite{b13}.

\subsection{Motivation for the Transition to 800 VDC Architecture}
Although traditional 48~VDC architectures have long enabled data center power scaling, several practical constraints emerge when supporting rack power beyond the 120-kW class, as highlighted below:
\begin{enumerate}
\item \textit{Excessive power loss and current limit}: At the rack level, low-voltage 48~VDC distribution requires very high current to deliver increasing power, which leads to severe resistive losses and increased heat generation \cite{b13,b39}.
\item \textit{Physical integration and scalability constraints}: Higher rack power often requires more and/or larger rack power inlets, which consumes rack space and complicates cable management \cite{b6}. Managing and protecting multiple AC feeds adds design complexity and increases the equipment footprint \cite{b6,b40}.
\end{enumerate}

800~VDC rack-level distribution has been proposed to address these challenges and provide several advantages, as highlighted below:
\begin{enumerate}
\item \textit{Fewer conversion stages and improved efficiency}: By converting grid power directly to 800~VDC at the data center entrance, intermediate conversion stages can be eliminated, and overall efficiency can be increased by up to 5\% \cite{b37}. By removing unnecessary stages, a leaner and more efficient structure is achieved, with fewer equipment layers, reduced cost and space, improved reliability, and faster deployment \cite{b13,b16,b41}.
\item \textit{Reduced copper usage}: By increasing distribution voltage, lower current is required for the same power, enabling thinner cables and reducing copper usage by up to 45\%, while also lowering cable weight and volume and improving cooling efficiency \cite{b37,b39}.
\item \textit{Simpler rack interface}: The power interface to IT can be simplified, since an 800~VDC approach can enable a more streamlined physical interface to the rack and can markedly reduce the number of cables and components \cite{b5,b16}.
\item \textit{Support for extreme rack power density}: Direct compatibility with high-power GPU racks is supported, because a higher rack busbar voltage can reduce conversion and routing volume in the compute space \cite{b16}.
\end{enumerate}

\subsection{Structure of the 800 VDC Architecture}
The 800~VDC architecture schematic is illustrated in Fig.~\ref{fig:2} \cite{b6,b13,b42}. In this schematic, a medium-voltage (MV) rectifier or a solid-state transformer (SST) is used as the front-end interface for converting medium-voltage AC (MVAC) to the 800~VDC bus. The MV rectifier and the SST are intended to enable direct conversion from MV levels (e.g., 35~kVAC) to 800~VDC, which can remove the need for the conventional low-voltage distribution layer (e.g., 480~VAC) and simplify facility-level power delivery.

This architecture aims to bring 800~VDC directly into the rack and distribute it to the compute node \cite{b16}. Inside the rack, the PSU first converts 800~VDC to an intermediate bus voltage, which is 48~VDC in early generations \cite{b13}. The on-board VRMs then step this IBV down to the 0.x~V (less than 1~V) rails required by the GPU domains. In this architecture, the DC--DC conversion stages are pushed as close as possible to the GPU to take advantage of the 800~VDC bus, minimizing power overhead in the system and reducing copper volume \cite{b43}.
Reference~\cite{b44} states that, for future AI data center designs, it is desired to reduce the intermediate bus voltage to levels below 48~VDC, to 12.5~VDC or even 6.25~VDC, to improve efficiency, simplify point-of-load regulation, and reduce power-delivery complexity. For a design in which the intermediate bus voltage is assumed to be 12.5~VDC, direct conversion from 800~VDC to 12.5~VDC is performed, and the 12.5~V rail is routed only a short distance to the VRMs. This direct step-down can be implemented using a 64:1 converter, eliminating the separate 800~VDC to 48~VDC and 48~VDC to 12.5~VDC stages. Compared with a multistage approach, this direct conversion has been reported to provide approximately 1\% additional efficiency improvement and about 26\% area reduction, which is particularly valuable near the GPU where space is most constrained \cite{b6}.

Another consequence of the 800~VDC architecture is the use of high-voltage DC (HVDC) servers. The effect of this change is seen at the latest stages of the power conversion chain, where the PSU, along with the VRM, is located in the HVDC server. Since the intermediate bus is brought close to the GPU and is itself being lowered, the two-stage VRM that converts the IBV to 0.x~V can be replaced by a single-stage VRM that converts the IBV directly to the core voltage \cite{b20,b45}. This VRM can be implemented either using the LPD or the VPD approach. These two approaches differ in several aspects, primarily due to their different effects on the PDN.

\section{Power Delivery Network}
\label{sec:pdn}
\begin{figure}[!t]
\centerline{\includegraphics[width=\columnwidth]{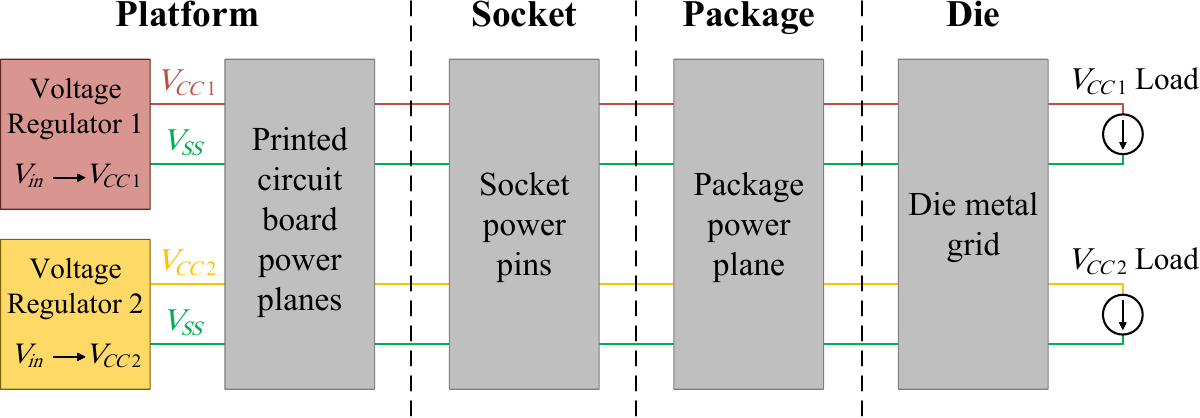}}
\caption{Elements of a typical PDN, from the platform to the die.\label{fig:3}}
\end{figure}

The power that leaves the VRM has to reach the transistors of the system on chip (SoC) die, and the network that carries it there is the PDN \cite{b20}. As shown in Fig.~\ref{fig:3}, this network contains every element between the voltage regulator output and the die, namely the power planes of the board on the platform, the power pins of the socket, the power planes of the package, and finally the metal grid of the die itself \cite{b46,b20}. The goal of the PDN is to deliver the current demanded by the load at every instant while holding the voltage at the die inside an allowable band. As the load current grows and its transients become faster, meeting this goal becomes one of the main design problems of AI servers, and the losses and parasitics of the PDN move to the center of attention \cite{b27,b28,b29}.

\subsection{Load Behavior and the Voltage Tolerance Band}
The computational core of an AI server is a high-performance processor built from smaller microprocessor units such as GPUs and central processing units (CPUs) \cite{b47}. These devices generally use flip-chip organic packages, in which a thick dielectric core provides the mechanical strength and the layers built above and below it carry the transistors together with the power and signal paths \cite{b20}. Every time a transistor is commanded to switch, a parasitic capacitance is energized by the power supply or de-energized by the load, and the stored energy is eventually dissipated as heat \cite{b20}. Since the transistors of an integrated circuit draw a DC component as well as an AC component, loss is produced through both mechanisms at the same time \cite{b20}. The AC component is the more demanding of the two, since the current drawn by the GPU can change at a rate of 2,000~A/$\mu$s or more when the device moves from the idle state to a fully loaded state \cite{b24}. To keep the load current as low as the required performance allows, dynamic voltage and frequency scaling (DVFS) is used in the microprocessor \cite{b48,b49,b50}.

Ideally, each transistor would receive exactly the voltage it needs at the instant it switches, but this cannot be achieved in practice. As a rule of thumb, a 5--10\% tolerance bound around the nominal voltage is therefore adopted as the allowable overshoot and undershoot \cite{b24,b20}. This band is the specification that the PDN has to meet, and the rest of this section describes the network through which it is met.

\subsection{Structure and Elements of the PDN}
The VRM stage is the first element of the PDN of an AI processor, and its input voltage in a data center server is usually 48~VDC \cite{b46,b20}. It converts and regulates the input voltage down to a core voltage below 1~V, and the resulting current flows through the board, the socket, and the package up to the load \cite{b51,b52}. In present designs, this conversion is usually carried out in two stages, in which the 48~V input is first brought to an intermediate 12~V rail and then to the core voltage. In new AI data center implementations, and also in the future 800~VDC power delivery architecture, this two-stage VRM is expected to be replaced by a single-stage VRM that converts the 48~V input directly to the core voltage \cite{b20}. Owing to the size and the layers of the printed circuit board (PCB), several power planes can be routed within a single board \cite{b20}. At the socket, several power and ground pins carry the current in parallel so that the parasitic impedance of the socket pins is kept low, and the number of pins assigned to a rail grows with the maximum current that the rail is expected to deliver \cite{b20}. This path is represented by a set of resistances, which govern the regulation, and inductances, which govern the transient behavior of the PDN \cite{b53}.

\begin{figure}[!t]
\centerline{\includegraphics[width=\columnwidth]{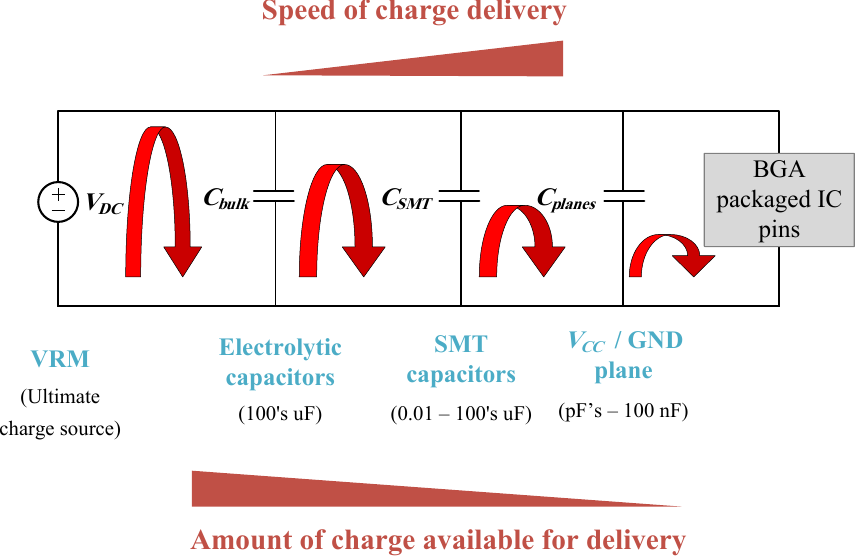}}
\caption{PDN charging and discharging hierarchy.\label{fig:4}}
\end{figure}

The layered construction of the PDN follows from a difference in time scales. The time in which a transistor at the load is commanded to change its state is a fraction of the time required by the VRM to respond with a change of its state \cite{b20}. The PDN is therefore represented as a hierarchy of charge sources, in which each stage delivers its charge faster than the one before it but holds a smaller amount of it, as illustrated in Fig.~\ref{fig:4} \cite{b20,b53,b54}. The VRM is the ultimate source of charge and the largest one, but it is also the slowest to respond. The electrolytic bulk capacitors that follow release their charge faster than the VRM, although the amount they can supply is smaller. The surface-mounted decoupling capacitors that come next are faster still and hold less charge again. At the end of the hierarchy, the power and ground planes supply the fastest charge of all, and the amount they hold is the smallest. Each of these stages is separated from the next by the interconnects modeled by an inductance and resistance that join them. Each of these capacitive, resistive, and inductive elements contributes to the impedance of the power delivery network.

\subsection{PDN Impedance and the Target Impedance Mask}
The losses of the PDN are expressed through its impedance, and power integrity design is therefore carried out in the frequency domain \cite{b33,b55}. Every capacitor component and every plane contribute a capacitive impedance, and every parasitic inductance contributes an inductive impedance that is proportional to its inductance \cite{b53,b56}. The components and the interconnects add their parasitic resistance as well \cite{b29}. Among these contributions, the current loop formed between the VRM and the load deserves particular attention \cite{b20}, and it is the reason why the high-power platform regulators, such as those feeding the cores, are usually placed as close to the microprocessor as possible \cite{b20}. This placement is not always achievable, however, since system-level constraints such as the breakout of the high-speed signals or the socket retention mechanism compete for the same space \cite{b20}.

\begin{figure}[!t]
\centerline{\includegraphics[width=\columnwidth]{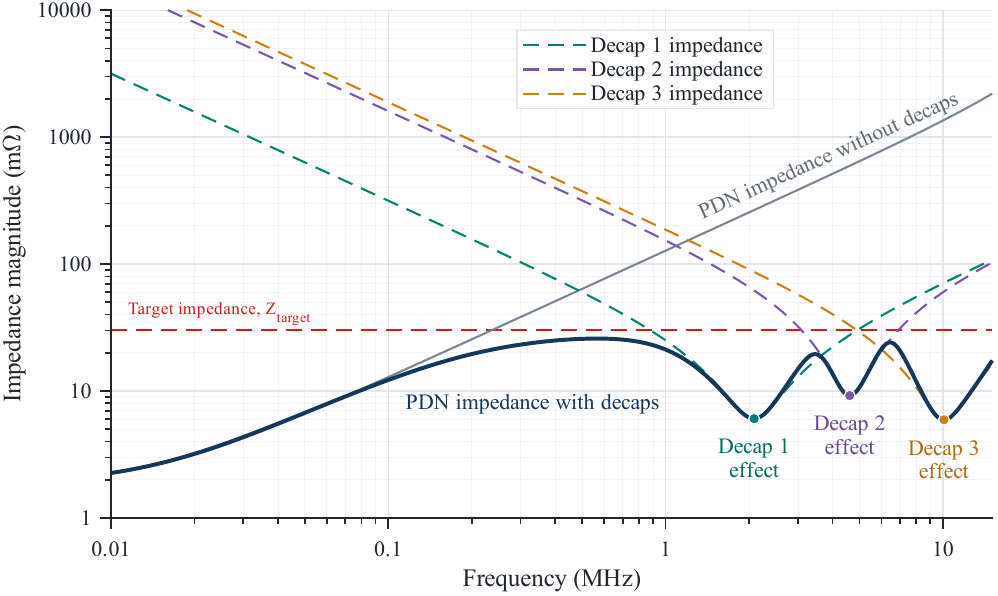}}
\caption{Typical optimized PDN resulting from assigning decoupling capacitors.\label{fig:5}}
\end{figure}

A high PDN parasitic impedance affects the delivered power to the load, where if the impedance is too large, the VRM is no longer able to supply the transient current that the load demands \cite{b53}. The current drawn by the load produces an instantaneous drop of voltage \cite{b20}, and the amplitude of this sudden transient response is set by the input impedance that the PDN shapes \cite{b33}. For this reason, a maximum impedance that must not be exceeded over the frequency range of interest is defined, known as the target impedance mask \cite{b33,b55}. This target impedance mask is obtained by dividing the allowable voltage droop by the largest load step that can occur during a transient \cite{b53},
\begin{equation}
Z_{\mathrm{target}}(f)=\frac{\Delta V_{\mathrm{droop}}}{\Delta I_{\mathrm{step}}}
\label{eq:ztarget}
\end{equation}
where $\Delta V_{\mathrm{droop}}$ is the allowable voltage droop and $\Delta I_{\mathrm{step}}$ is the maximum load step. Wherever the PDN impedance profile exceeds the mask, decoupling capacitors are added in the corresponding frequency range so that the impedance is brought back below the mask threshold \cite{b33,b55,b57}. This procedure is shown in Fig.~\ref{fig:5} \cite{b55}. The quality of a PDN can then be judged by how well its impedance profile stays below the mask from DC up to several hundred megahertz, where a low impedance over this range corresponds to an efficient power delivery network \cite{b20}.

\begin{figure}[!t]
\centerline{\includegraphics[width=\columnwidth]{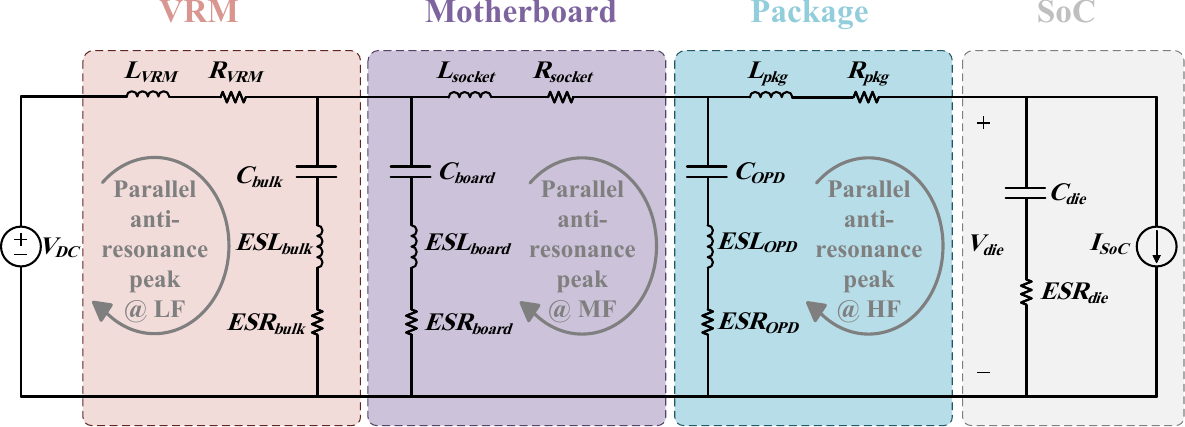}}
\caption{Circuit representation of a PDN with the three resonance peaks at low, mid, and high frequencies.\label{fig:6}}
\end{figure}

\subsection{Resonant Peaks and Voltage Droops}
\begin{figure}[!t]
\centerline{\includegraphics[width=\columnwidth]{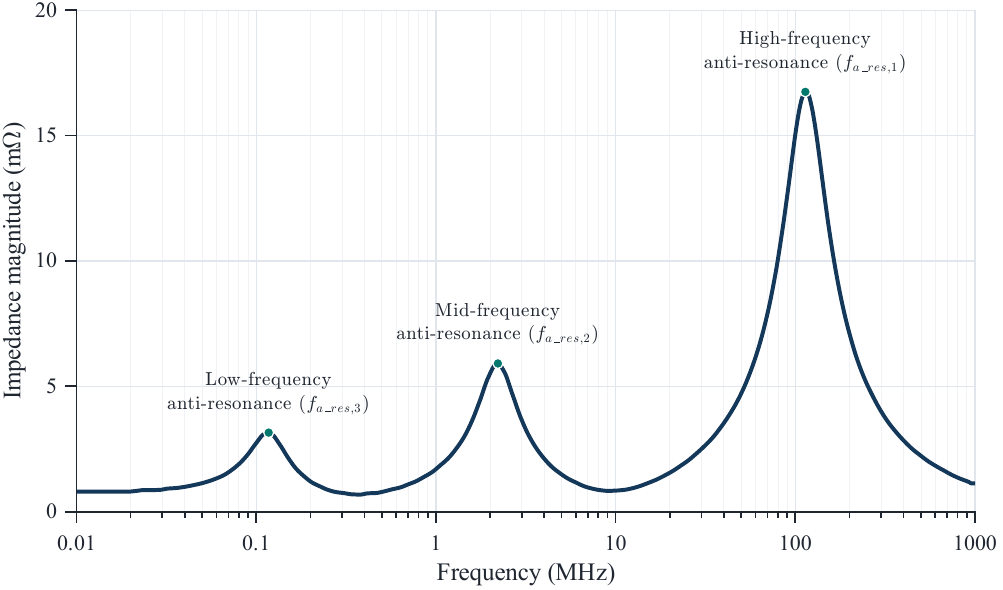}}
\caption{Impedance profile of a typical PDN in the frequency domain.\label{fig:7}}
\end{figure}

The lumped model circuit representation of a PDN is shown in Fig.~\ref{fig:6}. The interaction between its capacitive and inductive elements produces resonant frequencies, which appear as dips in the profile, and anti-resonant frequencies, which appear as peaks \cite{b33,b51,b55}. A typical PDN shows three anti-resonant peaks, and two resonant dips, whose positions are set by the parasitic capacitances and inductances of the network. The resonant frequencies are:
\begin{align}
f_{\mathrm{res},1}&=\frac{1}{2\pi\sqrt{\mathit{ESL}_{\mathrm{OPD}}\,C_{\mathrm{OPD}}}} \label{eq:fres1}\\
f_{\mathrm{res},2}&=\frac{1}{2\pi\sqrt{\mathit{ESL}_{\mathrm{bulk}}\,C_{\mathrm{bulk}}}} \label{eq:fres2}
\end{align}
And the anti-resonant frequencies are:
\begin{align}
f_{\mathrm{a\_res},1}&=\frac{1}{2\pi\sqrt{\left(L_{\mathrm{pkg}}\!+\!\mathit{ESL}_{\mathrm{OPD}}\right)\!C_{\mathrm{die}}}} \label{eq:fares1}\\
f_{\mathrm{a\_res},2}&=\frac{1}{2\pi\sqrt{\left(L_{\mathrm{socket}}\!+\!\mathit{ESL}_{\mathrm{board}}\!+\!\mathit{ESL}_{\mathrm{OPD}}\right)\!C_{\mathrm{OPD}}}} \label{eq:fares2}\\
f_{\mathrm{a\_res},3}&=\frac{1}{2\pi\sqrt{\left(L_{\mathrm{VRM}}\!+\!\mathit{ESL}_{\mathrm{bulk}}\right)\!C_{\mathrm{bulk}}}} \label{eq:fares3}
\end{align}
Here, $L_{\mathrm{pkg}}$, $L_{\mathrm{socket}}$, and $L_{\mathrm{VRM}}$ are the inductances of the package, of the socket connections, and of the VRM. $\mathit{ESL}_{\mathrm{OPD}}$, $\mathit{ESL}_{\mathrm{board}}$, and $\mathit{ESL}_{\mathrm{bulk}}$ are the effective series inductances (ESL) of the on-package decoupling capacitor, of the motherboard, and of the bulk capacitors. $C_{\mathrm{die}}$, $C_{\mathrm{OPD}}$, and $C_{\mathrm{bulk}}$ are the capacitances of the die, of the on-package decoupling capacitor, and of the bulk capacitor, respectively. The resulting impedance profile of a typical PDN in frequency domain is shown in Fig.~\ref{fig:7} \cite{b20}.

\begin{figure}[!t]
\centerline{\includegraphics[width=\columnwidth]{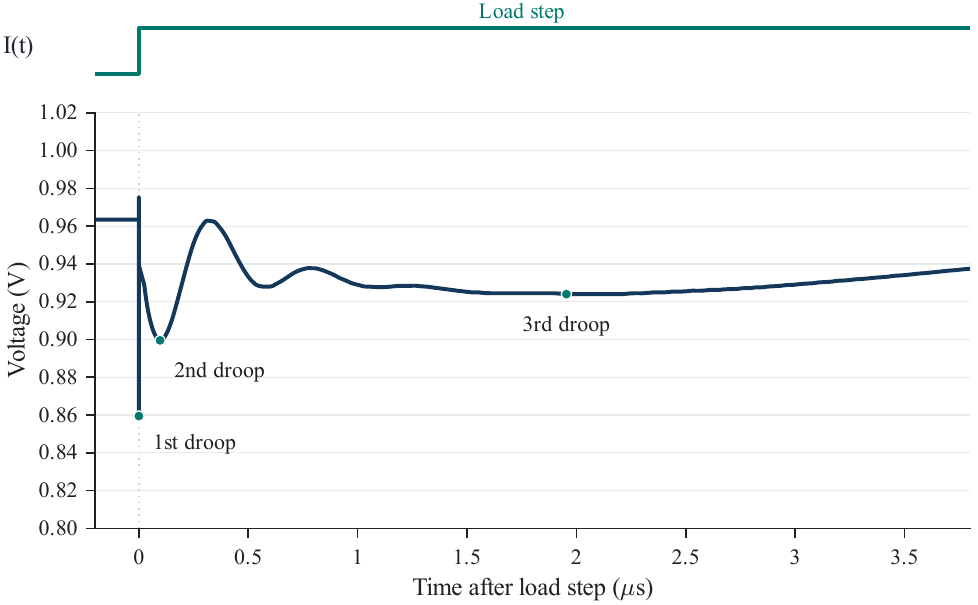}}
\caption{Time-domain voltage droops in a typical PDN.\label{fig:8}}
\end{figure}

The frequency domain is an intermediate representation of PDN behavior. In the time domain, the same peaks appear as rises and droops of the voltage while the die changes its state \cite{b20}. The first, second, and third peak produce the first, second, and third voltage droop, respectively, as shown in Fig.~\ref{fig:8} \cite{b20}. Since several droops occur one after another, the network becomes exposed to the rogue wave effect, in which the slower and smaller transients add to the faster and larger ones and create an impedance high point that rises above the target mask \cite{b20,b33,b58}. Two mechanisms keep this behavior under control. The resistance of the copper bumps, power and ground planes, interconnecting vias, and of the decoupling capacitors dampens the anti-resonant peaks and the ringing that accompanies the change of voltage \cite{b20,b50,b51}. In discrete systems, the two lower anti-resonance frequencies are handled by the decoupling capacitors. These capacitors are the element through which the designer has direct partial control over the transient response \cite{b20}.

\subsection{Decoupling Capacitor Selection and Types}
Decoupling capacitors are placed on the PCB, the package, and the die to compensate for the transients caused by the sudden current demand of the SoC \cite{b53,b55,b59}. Several stages of them are normally used so that current steps of different speeds can be served \cite{b20}. Each capacitor has an ESL and an effective series resistance (ESR), and these effects are represented in the lumped model by an inductor and a resistor \cite{b20,b53}. Two limitations have to be kept in mind. First, a capacitor does not necessarily stay at its rated capacitance when it is biased or when its temperature changes \cite{b20}. Second, a capacitor whose ESL is higher than that of the VRM ($L_{\mathrm{VRM}}$) would raise the inductance seen by the network, so only capacitors with a lower ESL should be used \cite{b53}. The choice is further constrained by the spread of the capacitance values, the parasitics of each candidate, and the cost \cite{b20}.

Adding decoupling capacitors also has consequences of its own. A larger number of capacitors pushes the anti-resonance frequencies of the PDN toward a higher frequency, which is often desirable \cite{b55}, but it also reduces the bandwidth of the PDN \cite{b60}. The opposite direction is equally effective, since a reduction of the inductance of the network, whether leakage, loop, or package inductance, lowers the decoupling capacitance that is needed to obtain a flat impedance profile \cite{b33,b52}. A systematic procedure for keeping the number of capacitors as low as possible is proposed in \cite{b55}, and it consists of four steps.
\begin{enumerate}
\item \textit{Selection of the target frequency}: A target frequency is chosen at the point where the impedance of the interconnects breaks through the mask \cite{b55}. Each time a capacitor is added, this target is updated so that the covered frequency range is broadened until the desired frequency is obtained \cite{b55}.
\item \textit{Screening of the candidates}: The ESR of each candidate capacitor is divided by the target impedance of the mask \cite{b55}. The result gives the number of capacitors of that ESR that would be needed to reach the target frequency, and the candidates that require the smallest number are kept \cite{b55}.
\item \textit{Ranking of the candidates}: Starting from the candidates that require the fewest capacitors, those that extend the frequency range the most are selected, whether they create anti-resonant peaks above the mask or not \cite{b55}.
\item \textit{Distribution of the values}: The final decision is whether the selected capacitors should all have the same value or should be a combination of different values \cite{b53,b55}.
\end{enumerate}
These steps are repeated until the target frequency is reached or exceeded, which makes the optimization of the capacitors relatively fast \cite{b55}.

\begin{figure}[!t]
\centerline{\includegraphics[width=\columnwidth]{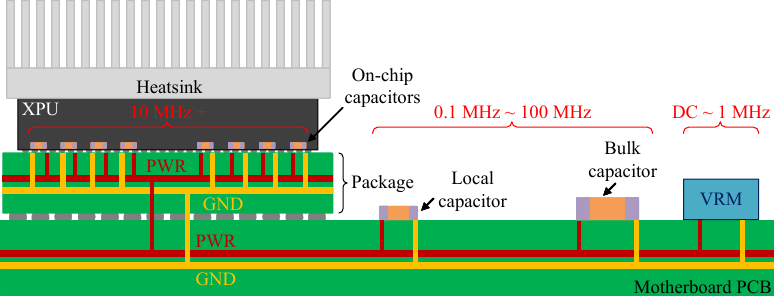}}
\caption{Placement of the decoupling capacitors along the PDN and the frequency range covered by each stage.\label{fig:9}}
\end{figure}

Several types of decoupling capacitor are combined to cover the wide frequency range of a PDN \cite{b55,b61,b62}. These capacitors are placed at different points along the path from the VRM to the die, and each of them covers a different part of the spectrum, as shown in Fig.~\ref{fig:9}. The VRM itself regulates from DC to the VRM switching frequency, which is usually about 1~MHz nowadays. The bulk and local capacitors mounted on the board cover the range between about 0.1~MHz and 100~MHz, and the on-chip capacitors take over above about 10~MHz. The bands overlap so that no part of the spectrum is left without a source of charge close enough to serve it. Table~\ref{tab:caps} shows some common capacitors used as decoupling capacitors and compares them in different aspects \cite{b20}. Electrolytic capacitors are usually the choice as the first decoupling stage after the VRM where height is not a constraint, since a high amount of charge must be available at this stage, while a high delivery speed is not yet required \cite{b20}. Tantalum polymer capacitors take less height and are therefore attractive where size matters, which is usually the case in data centers, but they are more expensive \cite{b20}. Multilayer ceramic capacitors (MLCCs), and in particular reverse geometry capacitors (RGCs) and interdigitated capacitors (IDCs), are the most widely used decoupling capacitors on the motherboard and on the package, owing to their high density and to their flexibility in shape \cite{b20}. The terminal augmentation of the RGC gives it a low ESR and a low ESL, and the use of several power and ground pins gives the IDC values that are even lower \cite{b20}. When an MLCC is mounted on the package, the shortened path to the load reduces its ESL further, and because of their high capacitance and their flexible size, MLCCs are normally placed on the top or on the bottom of the microprocessor package \cite{b20}.

\begin{table*}[!t]
\caption{Capacitor Types and Their Characteristic Properties}
\label{tab:caps}
\setlength{\tabcolsep}{3pt}
\renewcommand{\arraystretch}{1.25}
\begin{tabular*}{\textwidth}{@{\extracolsep{\fill}}lllllll}
\hline\noalign{\vskip 2pt}
\tabhead{Capacitor type} & \tabhead{Sub-classification} & \tabhead{Form factor} & \tabhead{Height class} & \tabhead{Capacitance} & \tabhead{ESL} & \tabhead{ESR / time constant} \\
\noalign{\vskip 2pt}\hline
\textbf{Electrolytic capacitors} & Aluminum capacitors & 5~mm diameter & 9~mm, 12~mm & 560~$\mu$F, 820~$\mu$F & 1.5--2.0~nH & 4--6~m$\Omega$ \\
 & & 10~mm diameter & 10~mm, 21~mm & 820~$\mu$F, 1.5~mF & 3--6~nH & 10--20~m$\Omega$ \\
 & Tantalum capacitors & 3528 & 1.1~mm, 1.9~mm & 220~$\mu$F, 470~$\mu$F & 500--1000~pH & 5--10~m$\Omega$ \\
 & & 7343 & 1.9~mm, 2.8~mm & 330~$\mu$F, 680~$\mu$F & 500--1500~pH & 10--20~m$\Omega$ \\
\hline
\textbf{Ceramic capacitors} & 2-terminal capacitors & 01005 & 0.15~mm, 0.22~mm & 0.1~$\mu$F, 0.22~$\mu$F & 120--150~pH & 4--6~m$\Omega$ \\
 & & 0201 & 0.15~mm, 0.22~mm & 0.1~$\mu$F, 0.47~$\mu$F & 150--200~pH & 10--20~m$\Omega$ \\
 & & 0402 & 0.22~mm, 0.33~mm & 1.0~$\mu$F, 2.2~$\mu$F & 200--250~pH & 5--10~m$\Omega$ \\
 & & 0603 & 0.5~mm, 0.7~mm & 4.7~$\mu$F, 10~$\mu$F & 300--350~pH & 10--20~m$\Omega$ \\
 & & 0805 & 0.7~mm, 1.4~mm & 22~$\mu$F, 100~$\mu$F & 300--400~pH & 4--6~m$\Omega$ \\
 & RGC capacitors & 0204 & 0.22~mm, 0.33~mm & 0.22~$\mu$F, 1~$\mu$F & 80--90~pH & 4--6~m$\Omega$ \\
 & & 0306 & 0.33~mm, 0.5~mm & 1~$\mu$F, 2.2~$\mu$F & 80--100~pH & 10--20~m$\Omega$ \\
 & IDC capacitors & 0603 & 0.5~mm & 1~$\mu$F, 2.2~$\mu$F & 30--50~pH & 5--10~m$\Omega$ \\
 & & 0805 & 0.7~mm & 2.2~$\mu$F, 4.7~$\mu$F & 50--60~pH & 10--20~m$\Omega$ \\
\hline
\textbf{Silicon capacitors} & MOS capacitor & N/A & N/A & 1--3~nF/mm$^{2}$ & Negligible & $RC<250$~ps \\
 & MIM capacitor & N/A & N/A & 20--200~nF/mm$^{2}$ & Negligible & 250~ps $<RC<5$~ns \\
 & DTC & N/A & N/A & 300--1500~nF/mm$^{2}$ & Negligible & 2~ns $<RC<20$~ns \\
\hline
\multicolumn{7}{@{}p{\textwidth}@{}}{Adapted from \cite{b63}. IDC, interdigitated capacitor; MIM, metal--insulator--metal; DTC, deep trench capacitor.}
\end{tabular*}
\end{table*}

The usefulness of the motherboard capacitors ends at about 10~MHz, because the inductance of the board, the ESL of the capacitor, and the parasitic inductance of the socket become too large \cite{b20}. Above this frequency, silicon capacitors built into the metal-oxide-semiconductor (MOS) structure of the transistors become important \cite{b20}. Their capacitance, however, is set by the applied bias because of the depletion behavior of the channel \cite{b20}, which makes them suitable mainly for low voltage regulation. At the end of the network, the capacitance available inside the die itself can also be used to reduce the input impedance seen by the load \cite{b33}, although the ESR of the on-die capacitors has to be watched carefully, since it can make them ineffective at very high frequencies \cite{b20}.

The behavior of the network therefore changes with frequency. The VRM and the bulk capacitors govern the low-frequency range, the motherboard and package decoupling capacitors govern the intermediate range, and the package and die capacitances govern the highest frequencies. What separates these ranges is not the capacitors themselves, but the parasitic inductance that lies between them and the die, and this inductance is fixed by the distance the current has to travel. The position of the regulator with respect to the load therefore shapes the impedance profile of the whole network, and it is examined in the following sections.

\section{Lateral Power Delivery}
\label{sec:lpd}
\begin{figure}[!t]
\centerline{\includegraphics[width=\columnwidth]{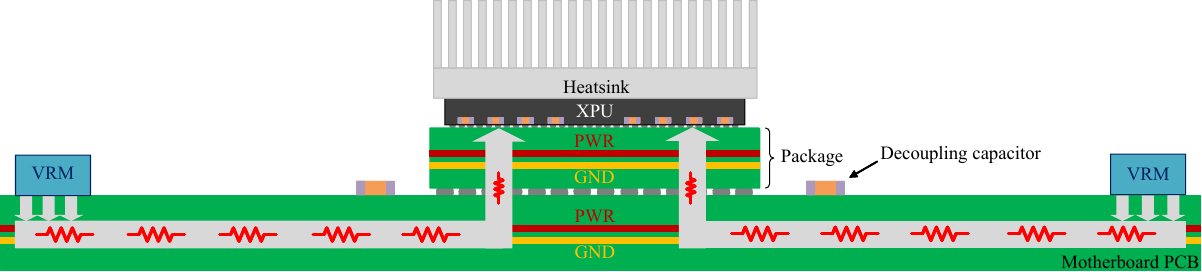}}
\caption{Lateral power delivery.\label{fig:10}}
\end{figure}

In a lateral PDN, the current flows laterally across the board, and the processor sits beside the VRMs that feed them, as illustrated in Fig.~\ref{fig:10} \cite{b26,b64,b65}. This arrangement has remained the dominant choice because it is cheaper than the alternatives, and its behavior is well understood \cite{b65,b66,b67}. Its one-sided construction leaves the other side of the board free, which gives easy access for cooling \cite{b60}. The power reaches the load through a chain of converters, capacitors, and connectors laterally \cite{b68}. In this section, LPD and its limitations that are moving the industry away from it are discussed.

The limitations of LPD fall into different aspects, including the voltage drop, increased loss and heat, and the competition for board area and pins. Each of them grows with the load current, and together they set the ceiling of the approach. These limitations are:
\begin{enumerate}
\item \textit{Voltage drops along the path}: The power loss across the PDN, from the regulators to the SoC die, is the main concern of LPD \cite{b20,b51}. The regulator adjusts its duty cycle so that the input voltage is converted to the desired output voltage \cite{b20}. The voltage of the die is the VRM output voltage minus the product of the load current and the total DC resistance of the path \cite{b20}. AI loads charge and discharge in bursts, and this behavior produces high spikes in the current they draw \cite{b69}. Whenever the load current changes suddenly, the voltage at the die follows, and a spike or a droop appears \cite{b20}. These transients degrade the performance and put the reliability of the microprocessor in question \cite{b20}. They can be reduced either by a faster regulator or by lowering the DC resistance as much as possible \cite{b20}.
\item \textit{Increased loss and heat}: The delivery path DC resistance turns part of the delivered power into heat, which is known as copper loss or Joule heating \cite{b20}. The long path of a traditional lateral PDN contains a large amount of copper, and this metal heats as the current passes through it \cite{b70,b71,b72}. At load currents beyond 1,000~A, even a small resistance measurably degrades the efficiency \cite{b67,b73,b74}.
\item \textit{Competition for board area}: A regulator placed laterally, away from the load, occupies board space that could otherwise be used for memory or for the fast communication links \cite{b67,b70,b73}. High bandwidth memory (HBM) already claims a considerable share of this space \cite{b63}. A group of decoupling capacitors are normally placed directly near the processor to handle the transients and to lower the PDN impedance \cite{b64,b75,b76}, and since they occupy the area near the die, the regulator is pushed to the side of it \cite{b53}. As the dies grow and the current density approaches and passes 1~A/mm$^{2}$ \cite{b77,b78}, the space left for the large passive components of the PDN shrinks further \cite{b43}.
\item \textit{Power pin count}: A lateral delivery scheme needs many power pins, which consume board area and complicate the design of the processor package \cite{b75,b79}. Each pin has a maximum current limit, so thousands of pins may be needed to carry the current \cite{b19}, and every pin assigned to power is taken away from the critical input/output (I/O) routing \cite{b77}.
\end{enumerate}

With LPD, the DC resistance can be lowered in several ways: by shortening the copper routing \cite{b30,b72}, by increasing the area, the thickness, and the number of the power planes, by using micro bumps and Cu--Cu hybrid bonding \cite{b72}, or by adding connection pins on the PCB and on the package \cite{b20,b63,b80}. Most of these measures, however, spend exactly the resource that is running out, since raising the performance of the system requires the area used for power delivery to become smaller and the PDN itself to become shorter \cite{b80}. The way out of this contradiction is not a better lateral network but a shorter PDN path, in which the delivery path is turned from the plane of the board into the vertical direction. This approach, VPD, is discussed in the next section.

\section{Vertical Power Delivery}
\label{sec:vpd}
\begin{figure}[!t]
\centerline{\includegraphics[width=\columnwidth]{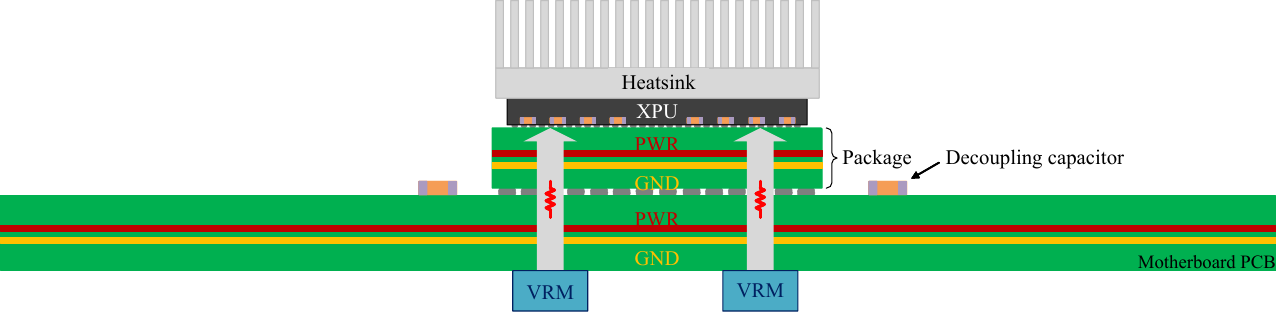}}
\caption{Vertical power delivery.\label{fig:11}}
\end{figure}

In VPD, the regulator is placed directly below the load, and the current flows vertically from there up into the SoC instead of laterally across the board, as illustrated in Fig.~\ref{fig:11} \cite{b33}. What turned this idea into an industry demand was the need to reduce the resistance and the length of the delivery path through the motherboard PCB \cite{b46,b78,b81}. Placing the conversion stage vertically shortens the copper path from the VRM to the die, which lowers both the resistance and the voltage loss \cite{b36,b82,b83}, and it frees the board area that the VRM used to occupy, which raises the power density \cite{b65,b77,b84}. The path resistance is reported to fall by up to a factor of 19 \cite{b63,b85,b86}, down to only 5--7~$\mu\Omega$ \cite{b27}. With such a low resistance, more current can be delivered from the regulator to the load without a higher loss, and the smaller voltage drop reduces the number of output capacitors that are needed \cite{b46}. This section describes the ways in which VPD is implemented, its advantages over LPD, the bottlenecks that still limit it, the designs reported so far, and the packaging technologies that make the vertical path possible.

\subsection{Implementation Approaches}
VPD can be implemented in two broad ways \cite{b12,b87,b88,b89,b90}. In the traditional form, the VRM, or several VRMs \cite{b72}, is mounted on the backside of the motherboard, directly below the die wherever this is possible \cite{b74,b91,b92}. Such a backside regulator can be built as a one-stage or a two-stage conversion \cite{b43,b47,b93}. The single-stage form removes the small loss of one of the converters and the need for the DC bus capacitors between the stages \cite{b43,b47,b93}. The regulator may also be mounted above the load, in what is called topside power delivery \cite{b94}, and in either case a one-sided arrangement remains relatively easy to cool \cite{b94}. Whether below or above the load, the power supply may need to be redesigned so that it can follow the needs of the microprocessor \cite{b46}. Delivery through the board is made possible by the blind and buried via process of through-silicon via (TSV) technology, which creates the vertical path while keeping most of the motherboard structure intact \cite{b46,b95}. The connections of the regulator board and of the motherboard may overlap, however, and if this is not accounted for, the vias can interfere with one another \cite{b46}.

The second broad approach integrates the regulator between boards or between dies, and it is the subject of the next section \cite{b12,b87,b88,b89,b90}. Between these two forms, a hybrid has also been proposed that combines vertical and lateral delivery into a power supply in package, which requires off-chip converters together with on-chip linear regulators \cite{b48}. This arrangement, sometimes called distributed VPD (DVPD), uses a two-stage conversion in which a low current is brought through the board to the neighborhood of the dies and the high current is produced by the second stage directly below the die \cite{b72,b96}. Whichever form is chosen, it has considerable advantages over LPD.

\subsection{Advantages Over LPD}
VPD is preferred over LPD because it raises the power and signal integrity, allows faster communication between processors, and increases the total power that can be delivered, all through the reduction of the PDN losses \cite{b97,b98,b99}. These gains come from three sources.
\begin{enumerate}
\item \textit{Lower path loss}: Shortening the current path lowers both the resistance and the inductance, and with them the power loss of the PDN \cite{b67,b100,b101}. Most of the remaining loss is then produced by the regulator itself, and the loss of the vertical path becomes almost negligible \cite{b19}. Shorter connections \cite{b30,b46,b33} also remove most of the parasitics that the motherboard used to introduce \cite{b101,b102,b103}.
\item \textit{Fewer decoupling capacitors}: A shorter path needs fewer decoupling capacitors \cite{b46}, and the microprocessor receives its power from the regulator faster. This matters because a regulator mounted on the backside of the motherboard takes some of the area once reserved for the output capacitors \cite{b60,b104}. If decoupling capacitors are still required, one direction under study is to embed them inside the regulator \cite{b95}. The number of input capacitors can be reduced as well, since an unregulated input stage removes the need for many of them and opens the possibility of placing a regulator even closer to the die \cite{b33,b105}.
\item \textit{Thermal relief and board area}: With less copper in the current path, less heat is produced, and the cooling system on top of the PCB is less stressed \cite{b76}. The freed board area can be used for more memory or for signal routing \cite{b67,b101,b104}. A high power density is required from the VRM in any case, since it must exceed that of the microprocessor in order to supply the instantaneous currents when they are demanded \cite{b106}.
\end{enumerate}

\begin{table*}[!t]
\caption{Summary of Reported Vertical Power Delivery Designs}
\label{tab:vpd}
\setlength{\tabcolsep}{2pt}
\renewcommand{\arraystretch}{1.25}
\begin{tabular*}{\textwidth}{@{\extracolsep{\fill}}clcccccccc}
\hline\noalign{\vskip 2pt}
\tabhead{Year} & \tabhead{Topology} & \tabhead{$V_{\mathrm{in}}/V_{\mathrm{out}}$\\(V)} & \tabhead{$I_{\mathrm{out}}$\\(A)} & \tabhead{$f_{\mathrm{sw}}$\\(kHz)} & \tabhead{Height\\(mm)} & \tabhead{Current\\density\\(A/mm$^{2}$)} & \tabhead{Power\\density\\(W/in$^{3}$)} & \tabhead{Power stage\\efficiency\\peak/full (\%)} & \tabhead{System\\efficiency\\peak/full (\%)} \\
\noalign{\vskip 2pt}\hline
2025 & SBC-VPD 8-branch \cite{b93} & 48/1 & 400 (25/phase) & 600 & 4.7 & 0.501 & 1745 & 90.2/85.1 & 88.2/84.4 \\
2025 & SBC-GaN \cite{b109} & 48/1 & 405 (25.3/phase) & 1000 & 2.7 & 0.502 & 3048 & 88.7/76.7 & 86.5/76.3 \\
2025 & Kong \textit{et al.} \cite{b46} & 12/1.83 & 220 & 600 & -- & -- & -- & -- & 93.0/91.5 \\
2025 & 2-D scalable planar DCX \cite{b28} & 58/1.8 & 450 & 1200 & 8 & 0.25 & 924 & 94.4/86.1 & -- \\
2025 & Dual-inductor SBC \cite{b110} & 48/1 & 50 (25/phase) & 1000 & 2.6 & 0.321 & 2020 & 91.9/88.0 & 90.1/87.3 \\
2024 & Planar matrix DCX \cite{b111} & 48/1 & 350 & 150 & 5 & 0.08 & 273 & 94.0/85.0 & -- \\
2024 & Ultra-low LLC-DCX \cite{b112} & 36/0.75 & 150 & 1000 & -- & 0.11 & 255 & 93.7/89.0 & -- \\
2024 & 20:1 SBC \cite{b47} & 48/1 & 1500 (37.5/phase) & 220 & 6.2 & 0.287 & 759 & 94.1/86.0 & 92.7/85.7 \\
2024 & 16:1 SBC \cite{b113} & 48/1 & 500 (31.3/phase) & 150 & 7.3 & 0.207 & 464 & 94.7/86.0 & 93.4/86.1 \\
2024 & Mini LEGO \cite{b114} & 48/1 & 240 (20/phase) & 1500 & 8.4 & 0.71 & 1390 & 87.1/84.1 & 84.1/82.3 \\
2024 & SDIH \cite{b115} & 48/1 & 105 (52.5/phase) & 750 & 3.6 & 0.145 & 598 & 83.5/71.5 & 81.4/70.9 \\
2023 & LLC-DCX \cite{b116} & 48/1 & 300 & 1000 & 5.6 & 0.34 & 1010 & 94.1/86.8 & -- \\
2023 & MSC \cite{b117} & 48/1 & 450 (28.1/phase) & 400 & 7 & 0.265 & 621 & 93.1/84.1 & 91.7/85.8 \\
2023 & SBC \cite{b118} & 48/1 & 1200 & 200 & -- & 0.23 & 607 & 92.4/87.5 & -- \\
2023 & MHB-CDR \cite{b119} & 48/1.8 & 120 & 600 & -- & 0.15 & 1037 & 93.1/87.8 & -- \\
2022 & Dickson square \cite{b120} & 48/1 & 270 (30/phase) & 280 & 6.45 & 0.142 & 360 & 93.8/88.4 & 91.6/87.7 \\
2022 & VIB \cite{b121} & 48/1 & 450 (28.1/phase) & 417 & 11 & 0.156 & 232 & 95.2/89.1 & 93.3/88.1 \\
2022 & MLB \cite{b122} & 48/1 & 60 (30/phase) & 250 & 7.8 & 0.125 & 263 & 92.7/88.6 & 91.5/88.4 \\
2022 & LEGO \cite{b123} & 48/1 & 450 (28.1/phase) & 1000 & 16.65 & 0.298 & 294 & 91.1/85.7 & 88.4/84.4 \\
2021 & 3-D LEGO PoL \cite{b124} & 48/1.5 & 450 & 1000 & 20.5 & 0.57 & 688 & -- & 93.3/87.2 \\
2020 & Cross-coupled QSD buck \cite{b125} & 48/1 & 40 (20/phase) & 125 & -- & -- & 150 & 95.1/92.7 & -- \\
2020 & Sigma \cite{b126} & 48/1 & 40 (20/phase) & 1000 & 4 & 0.127 & 420 & 94.0/86.0 & 92.5/85.7 \\
\hline
\end{tabular*}
\vspace{18pt}
\end{table*}

\begin{figure*}[!t]
\centering
\panel{0.48\textwidth}{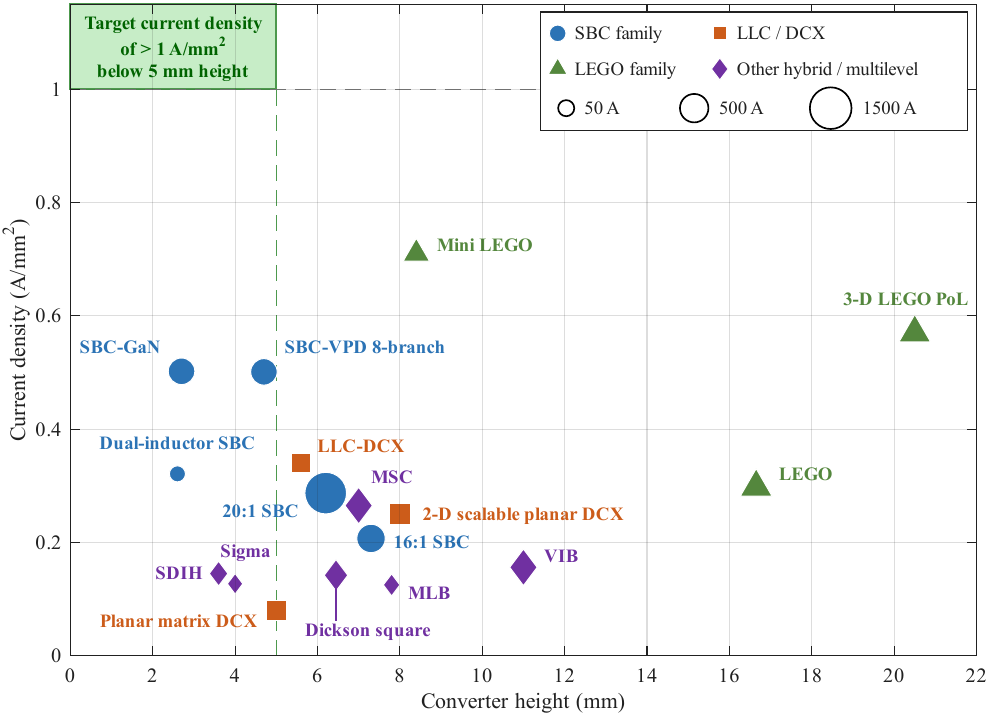}{a}\hfill
\panel{0.48\textwidth}{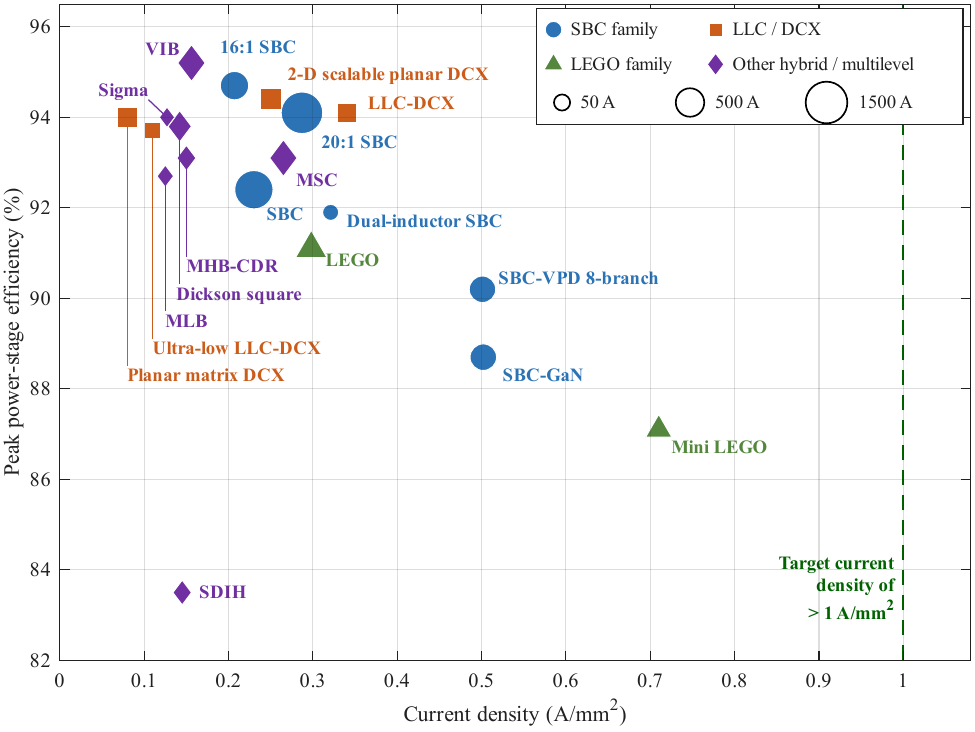}{b}
\caption{Design space of the VPD converters of Table~\ref{tab:vpd}. (a) Converter height versus current density. (b) Current density versus peak power-stage efficiency.\label{fig:12}}
\end{figure*}

\subsection{VPD Bottlenecks}
VPD removes the limitations of the lateral path, but it is not free of its own.
\begin{enumerate}
\item \textit{Height}: Keeping the height of the regulator in the low millimeter range is one of the strongest constraints on a smaller VRM for VPD \cite{b43,b107,b82}. Since the current is confined to the vertical direction, the inductors are limited not only in height \cite{b74,b101} but also to vertical designs for their windings and cores \cite{b82}.
\item \textit{Redistribution board}: A backside regulator may require a redistribution board, which brings back some lateral loss, adds rails, and may call for additional decoupling capacitors and further design effort \cite{b60}.
\item \textit{Thermal management}: The components are thermally coupled, and the height limit keeps shrinking, so the heat has to be removed through thermal vias, cold plates, or other means of cooling \cite{b104,b108}. Most cold plates are designed for top-side cooling, and only a few solutions exist for cooling the bottom side \cite{b43}.
\item \textit{Absence of a standard}: No standard has yet been established for VPD. Reference~\cite{b109} proposed that the VRM should reach a current density of 1~A/mm$^{2}$ at a height below 5~mm, and this target has still not been met.
\end{enumerate}

\subsection{Reported Designs}
Topologies such as series-input parallel-output converters are being studied for their ability to be scaled up, together with multilevel multiphase flying-capacitor converters, which are becoming faster and more reliable \cite{b106}. Table~\ref{tab:vpd} collects the VPD designs reported between 2020 and 2025, most of them for the conversion of 48~V to about 1~V, with their output current, switching frequency, height, current density, power density, and efficiency. Fig.~\ref{fig:12} maps these designs by height and current density in (a), together with the target of \cite{b127}, and by current density and peak efficiency in (b). The output currents range from 40~A to 1500~A, the switching frequencies from 125~kHz to 1.5~MHz, and the heights from 2.6~mm to 20.5~mm. The table also shows how far the target of \cite{b127} still is. The highest current density, 0.71~A/mm$^{2}$, is reached at a height of 8.4~mm \cite{b114}, and the lowest heights, 2.6--2.7~mm, are reached at 0.3--0.5~A/mm$^{2}$ \cite{b109,b110}, so that no design listed combines 1~A/mm$^{2}$ with a height below 5~mm.

\subsection{Packaging and Interconnect Technologies}
\begin{figure*}[!t]
\centerline{\includegraphics[width=\textwidth]{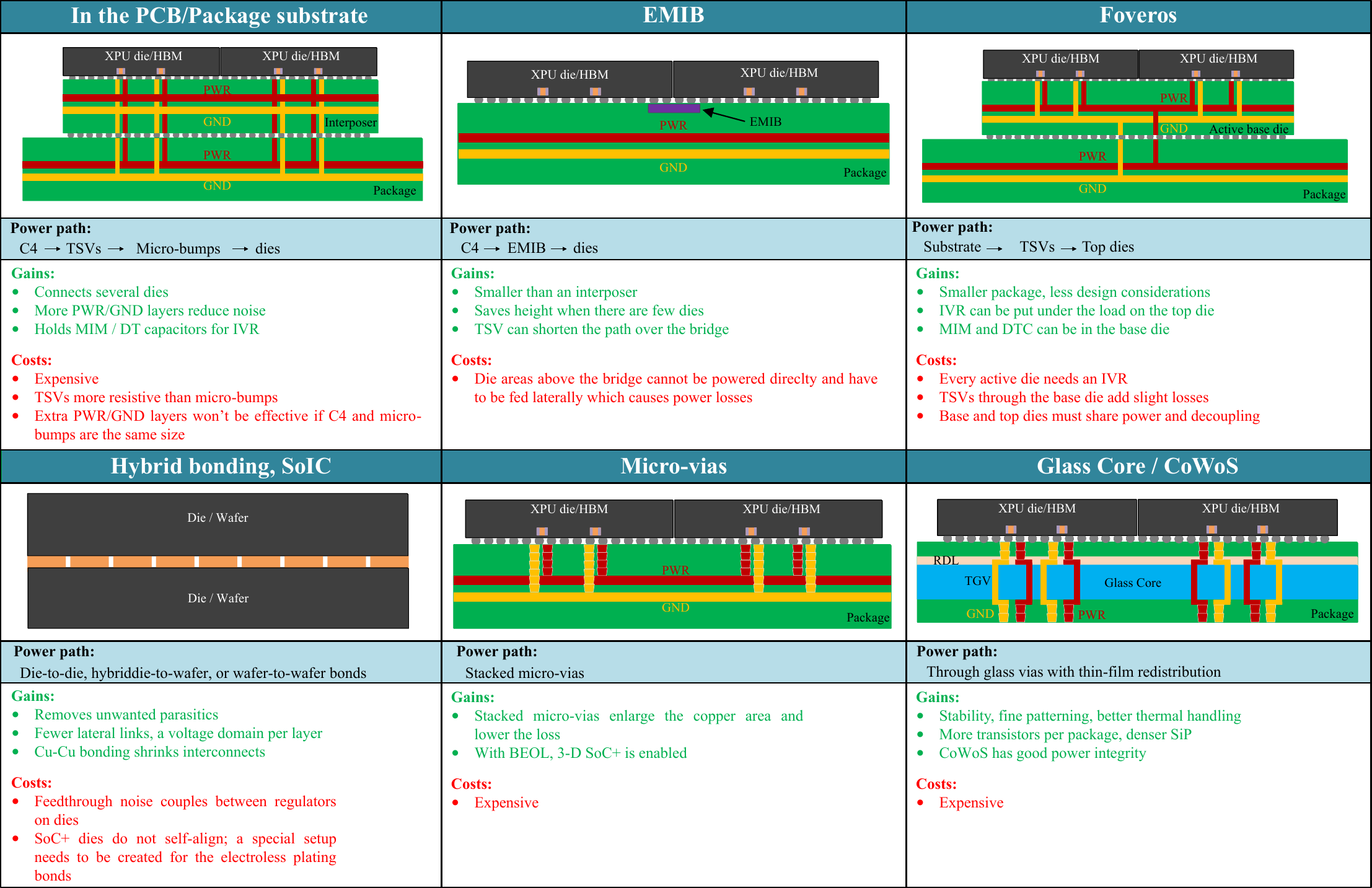}}
\caption{Packaging and interconnect technologies for VPD, with the power path, gains, and costs of each.\label{fig:13}}
\end{figure*}

The idea behind VPD is older than the term, and it began as an effort to save space by combining several circuits in one structure. Component embedding was introduced for ceramic circuits in the 1980s and reached practical use with organic PCBs in the 1990s \cite{b128}. Three-dimensional (3-D) stacked packages followed, driven by the need to fit more synchronous dynamic random-access memory (SDRAM) into cellphones \cite{b129}, and the wafers stacked in this way, connected by interchip vias (ICVs), were called vertically integrated circuits (VICs) \cite{b130}. To carry chip-to-chip packaging further, the TSV was introduced to connect several layers to one another \cite{b129}. The TSV lowers the signal energy per bit and the link latency while raising the connection density \cite{b26,b131}, and a prototype in \cite{b129} showed a low contact resistance together with an eye diagram that confirmed its accuracy up to 20~GHz. TSVs made die-to-die stacking possible, while controlled collapse chip connection (C4) bumps gave the dies their copper connections \cite{b132}, and the thicker the C4 bumps, the smaller the voltage drop across them \cite{b133}. The next step was to add more die stacks to the same package through an interposer \cite{b132}. This idea had been proposed in the 1960s, but it became feasible only once back-end die stacking was available \cite{b132}. The main packaging and interconnect options are summarized in Fig.~\ref{fig:13}, together with the power path, the gains, and the costs of each.

Technologies such as the silicon interposer \cite{b134}, the embedded multi-die interconnect bridge (EMIB) \cite{b135}, and Foveros \cite{b136} give VPD a way to increase the connection density \cite{b20}. These newer heterogeneous integration techniques are classified as 2.5-D or 3-D \cite{b20}. In 2.5-D packaging, two or more integrated circuits (ICs) are interconnected through the same package \cite{b20}. A silicon interposer can connect several dies, but it has to be large enough to hold all of them, which makes it expensive \cite{b20}. The dies are attached to the interposer through micro bumps, and the more power and ground layers the interposer contains, the better the noise is reduced \cite{b133}. If the C4 bumps below the interposer have the same size as the micro bumps above it, there is little reason to add power and ground layers, because the parasitics of the TSVs and of the micro bumps are what spread the current through the interposer \cite{b133}. The EMIB is a smaller solution that connects the chips through a bridge \cite{b20}. The parts of the dies that sit on the bridge cannot be powered directly and are instead fed laterally through the other copper bumps \cite{b20}, and with bridge technologies the TSVs can shorten the delivery path in the overlapping areas \cite{b133}.

In 3-D packaging, two or more ICs are stacked on top of one another, which improves the performance \cite{b80}, and the approach began as a way to increase memory by stacking \cite{b20}. A 3-D IC needs fewer lateral interconnections, and it supports heterogeneity in an SoC, since its layers can be made independently and stacked afterwards \cite{b80,b66}. For this heterogeneity to be effective, each layer is allowed its own voltage level, which raises the responsiveness by increasing the number of voltage domains \cite{b80}. The drawback of such a heterogeneous regulator arrangement is that feedthrough noise from one source couples to its load and from there affects the loads of the other regulators on the same die \cite{b33,b137}. The interconnects of a 3-D IC can be die-to-die, hybrid die-to-wafer, or monolithic wafer-to-wafer \cite{b75,b132}, and these integration techniques remove most of the parasitics between the interconnects \cite{b75}. Fine-pitch plated through holes (PTHs) also give the regulator a direct DC contact to the substrate \cite{b138}. An example of the density that can be reached is Intel's Ponte Vecchio processor, which is built in a 7~nm process from 47 tiles and is designed for 600~W \cite{b68,b86}.

Foveros resembles the original 3-D stacking used for memory blocks, but it stacks active dies on active dies instead of passive circuits \cite{b20,b139,b140}. Every die therefore has to be powered by an IVR, and the power for the top dies passes through the base die by TSVs, as in a silicon interposer, which raises the power loss slightly \cite{b20}. In return, the package becomes smaller and fewer die design considerations arise \cite{b20}. The question that Foveros raises is how to power the base die and the top dies at the same time, since both are active, and one answer is to share the power and the decoupling between them \cite{b20}. A related technology is the system on integrated chip (SoIC) \cite{b20,b140}. To raise the density of vertically stacked boards further, Cu--Cu direct bonding is being studied \cite{b131}, as it reduces the size of the usual ball grid arrays (BGAs) and C4 bumps \cite{b30}. A back-end-of-line (BEOL) approach combines the connectivity of TSVs and of 3-D ICs \cite{b131}, and high-density interconnects in the PCB, such as micro-vias, offer an even smaller connection \cite{b51}. Micro-vias can be stacked on one another, which enlarges the effective copper area compared with normal vias and lowers the power loss \cite{b51}. Together, BEOL and micro-via techniques allow a 3-D SoC+ design, in which several active chips are embedded in one die \cite{b131}. Its drawback is that the chips do not self-align, and \cite{b131} uses electroless plating bonds instead of Cu--Cu bonding, which adds complexity and requires a special setup.

Chip-on-wafer-on-substrate (CoWoS) has also been introduced as a system with good power integrity \cite{b57}. Even so, the reduction of the packaging resistance brought by the higher density has progressed slowly compared with the growth of the power loss to be handled \cite{b30}. More recently, glass substrates have been proposed in place of the organic core for their stability, their patterning capability, and their thermal handling, among other properties \cite{b59,b141,b142}. Glass had been examined years earlier but was set aside because of its high cost \cite{b132,b141}, and Intel has now revived it by launching its own glass packaging for AI and HPC systems \cite{b132}. The glass replaces the organic core while still relying on through holes, called through-glass vias (TGVs), with metal interconnects and thin-film dielectrics for the redistribution routing \cite{b132}. The capabilities of the glass substrate allow more transistors to be fitted into a package, which produces an even denser system in package (SiP) \cite{b142,b143}.

All of these technologies tighten the connection between the regulator and the die, and the vertical path they create is the one that the designs of Table~\ref{tab:vpd} exploit. Once the interconnect is this dense, the next step is no longer to place the regulator below the package but to integrate it within it. This step, the IVR, is examined in the following section.

\section{Integrated Voltage Regulators and Embedded Passive Components}
\label{sec:ivr}
\begin{figure}[!t]
\centerline{\includegraphics[width=\columnwidth]{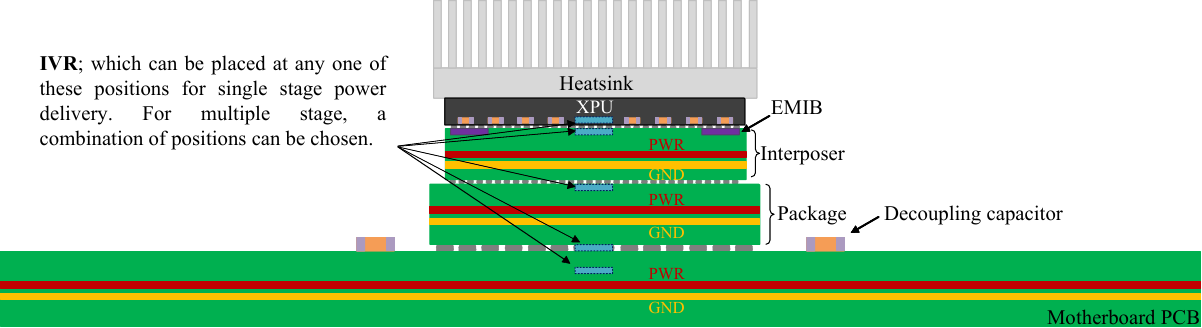}}
\caption{Possible IVR positions across the die, interposer, package, and motherboard PCB.\label{fig:14}}
\end{figure}

As VPD has matured, the regulator has been pushed step by step toward the load, from the backside of the motherboard to the space between the motherboard and the package, into the package substrate, onto the interposer, and finally onto the die itself \cite{b20,b33,b66}. The driving force behind this migration is the rapid growth in the number of on-die power domains together with the continued growth of the processor power \cite{b20}, and the use of integrated regulators in commercial products is already visible \cite{b31,b34,b144}. An IVR allows the passive components to be placed in the die, in the interposer, or in the package, which reduces the area and the complexity of the PCB \cite{b59,b78,b145}. As the passive components become smaller and are placed closer together, the area taken by the regulator shrinks and its power density rises \cite{b33,b77}, and the existing interconnect technologies can be kept while the transient response improves and the power loss falls \cite{b80,b146}. Possible IVR positions are illustrated in Fig.~\ref{fig:14}. This section follows the regulator along its path toward the die, from the choice of integration level and topology, through the fully integrated form, to the inductors, capacitors, and switches that have to be shrunk to make it possible, and finally to the thermal problem that this shrinking creates.

\subsection{Levels of Integration and Regulator Topologies}
\label{sec:ivr-levels}
The benefit of integration comes mainly from the input voltage. When an IVR is implemented correctly, the power enters the package at a higher voltage and a lower current, which reduces the resistive loss of the PDN even further than backside mounting does, and the conversion loss added by the extra stage is offset by operating with a larger duty cycle \cite{b30,b72,b83}. Hybrid regulators such as switched-capacitor (SC) converters are better suited to this higher voltage scaling than a plain buck converter, although their higher switching frequency makes them less efficient than fixed-ratio converters \cite{b30,b107,b147}. The requirements also differ between processor types. A CPU needs several small to medium voltage domains per core, its transient response has to be tracked, and its regulator can be integrated into the SoC \cite{b104}. A GPU needs large voltage domains and a well-controlled transient response, but per-core domains are not necessary and the regulator is not integrated into the SoC \cite{b104}. For high-ratio and multistage conversion, newer circuit topologies and wide-bandgap materials are worth examining, provided that the high efficiency is not compromised \cite{b30}.

The conversion ratio sets a hard limit on how small the regulator can become. A high-ratio regulator that brings 48~V down to 1~V has to operate with a very small duty cycle, which limits its switching frequency and the on-time of the high-side switch, rules out embedded inductors, and therefore limits the size reduction that can be reached \cite{b79}. The small duty cycle also raises the voltage spikes and the root mean square (RMS) current of the high-side switches, which increases the copper, inductor, and switching losses \cite{b79,b133,b148}. Topologies based on series capacitors have therefore been adopted so that a larger duty cycle can be used and the efficiency raised \cite{b79}. Other arrangements chain the regulators together or spread them in a distributed array so that the power is delivered uniformly \cite{b82}. Both require more I/O pins and more distribution lines across the boards for sensing and redundancy, but both allow several outputs to feed the load at the same time \cite{b82,b149}.

\begin{figure*}[!t]
\centerline{\includegraphics[width=\textwidth]{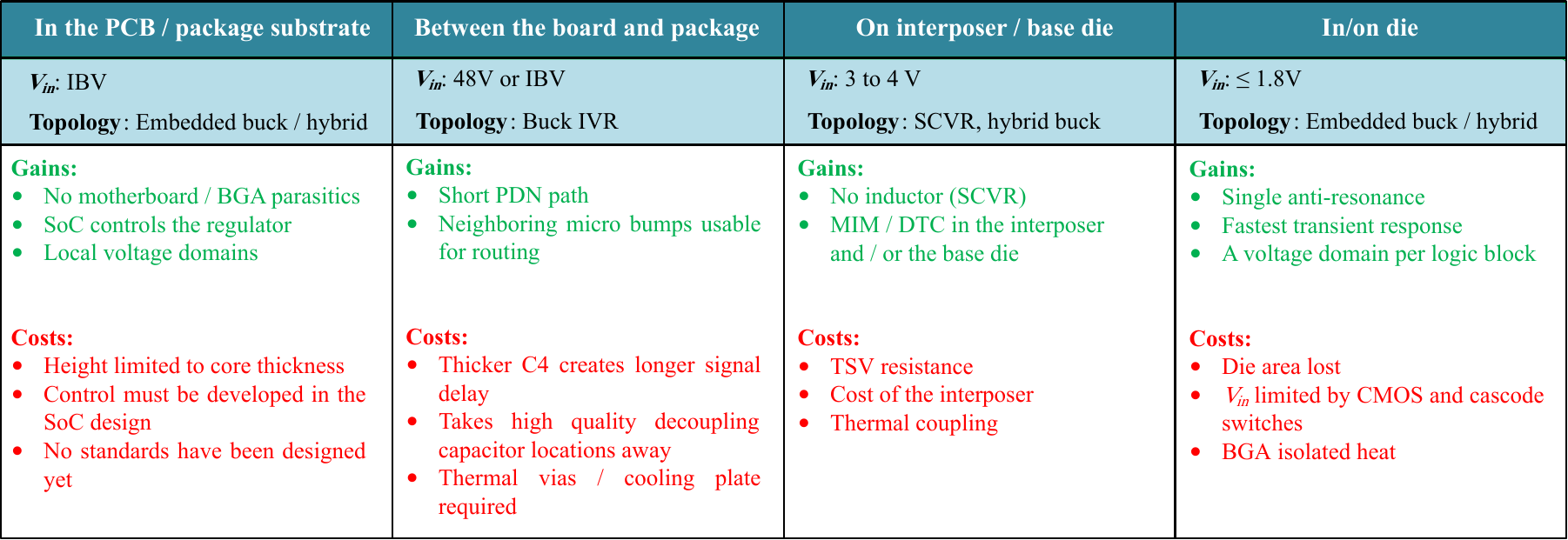}}
\caption{Levels of regulator integration and their trade-offs.\label{fig:15}}
\end{figure*}

Each level of integration brings its own trade-offs, which are summarized in Fig.~\ref{fig:15}. When the IVR is placed between the motherboard and the package, it is usually larger than the copper bumps that carry the power and the signals, so the bumps have to be made thicker for the regulator to fit \cite{b46}. Thicker bumps move the CPU or the GPU farther from the signals routed through the motherboard, which lengthens the delay and adds some interconnect resistance \cite{b46}. Placing the IVR below the die on the package takes away the best location for the decoupling capacitors and exposes the regulator to heating from the die \cite{b46}. If the regulator is placed between the die and the package in a low-current area, the neighboring copper bumps can be used for current routing \cite{b20,b114}, but removing parallel current paths raises the resistance and forces higher currents through the remaining bumps \cite{b107}. Cooling also becomes an issue, since vias are needed to spread the heat toward the cold plates, and a bottom cold plate may be required in addition \cite{b107}.

At the level of the base die, the switched-capacitor voltage regulator (SCVR) has emerged as the preferred candidate. A buck converter placed on the base die needs twice the number of pins to route the current to and from its package inductors, its input voltage is limited to about 1.8~V by the scaled complementary metal-oxide-semiconductor (CMOS) devices, and its thick-core package raises the assembly cost \cite{b150}. An SCVR built with on-die MIM capacitors needs no inductor and no intermediate switch-node connection to the package, and its flying capacitors lower the voltage stress on the switches so that a higher input voltage can be accepted \cite{b150}. Fixed-ratio designs give the best current density and efficiency for a given capacitor and device figure of merit, and the load and line regulation that they lack can be supplied by on-die low-dropout (LDO) regulators \cite{b150}. A monolithic multistage modular SCVR in 16~nm CMOS has reached a peak efficiency of 90.6\% at a 3:1 ratio, with a peak current density of 5.7~A/mm$^{2}$ from a 3~V input, and 87.3\% at a 4:1 ratio from a 4~V input, and its 35 modules deliver up to 7~A from a die area of about 1.23~mm$^{2}$ \cite{b150}. Table~\ref{tab:ivr} compares this design with the integrated regulators reported before it. The monolithic designs with on-die capacitors reach the highest current densities, whereas the designs that rely on package inductors or MLCCs remain below about 0.4~A/mm$^{2}$ \cite{b150}. The next step beyond the base die is to integrate the regulator completely, together with its passive components.

\begin{table*}[!t]
\caption{Comparison of Integrated Voltage Regulators}
\label{tab:ivr}
\setlength{\tabcolsep}{2pt}
\renewcommand{\arraystretch}{1.25}
\begin{tabular*}{\textwidth}{@{\extracolsep{\fill}}cccllcccccc}
\hline\noalign{\vskip 2pt}
\tabhead{Ref.} & \tabhead{Year} & \tabhead{Node\\(nm)} & \tabhead{Topology} & \tabhead{Passive elements} & \tabhead{Monolithic} & \tabhead{Conversion\\ratio} & \tabhead{$V_{\mathrm{in}}$\\(V)} & \tabhead{$V_{\mathrm{out}}$\\(V)} & \tabhead{Peak\\efficiency (\%)} & \tabhead{Max. current density,\\passives included (A/mm$^{2}$)} \\
\noalign{\vskip 2pt}\hline
\cite{b44} & 2022 & 22 & Buck & In-package inductor & No & 1.5:1 & 1.8 & 1.2 & 85.2 & 0.1 \\
\cite{b150} & 2025 & 16 & MMSCVR & On-die MIM & Yes & 3:1 & 3 & 1 & 90.6 & 5.67 \\
\cite{b150} & 2025 & 16 & MMSCVR & On-die MIM & Yes & 4:1 & 4 & 1 & 87.3 & 2.96 \\
\cite{b151} & 2017 & 130 & Hybrid buck & MLCC & No & 1.83:1 & 3.3 & 1.8 & 88 & 0.3 \\
\cite{b152} & 2024 & 65 & Hybrid buck & MLCC & No & 3.6:1 & 3.6 & 1 & 93.9 & 0.27 \\
\cite{b153} & 2023 & 65 & SCVR & MLCC & No & 4:1 & 4.2 & 1 & 92.5 & 0.38 \\
\cite{b154} & 2023 & 4 & CSCR SCVR & On-die MIM & Yes & Continuous & 1.3 & 0.95 & 88.5 & 2.2 \\
\cite{b155} & 2017 & 28 & SCVR & On-die MOM + MOS & Yes & 3:1 & 3.2 & 0.95 & 82 & 1.16 \\
\cite{b156} & 2014 & 32 & SCVR & Deep trench capacitor & Yes & 3:2 & 1.8 & 1.1 & 90 & 3.37 \\
\hline
\multicolumn{11}{@{}p{\textwidth}@{}}{MOM, metal--oxide--metal.}
\end{tabular*}
\end{table*}

\subsection{Fully Integrated Voltage Regulators}
When the regulator and all of its passive components are integrated together and placed inside or between the components of the package, the result is referred to as a FIVR \cite{b80,b66}. A FIVR makes it possible to provide many different voltage domains and requirements, but if it is integrated on the die it takes away valuable die area \cite{b78,b80}. The first FIVR reported in \cite{b104} used an air-core inductor and reached an efficiency of 90\% when converting 1.8~V to 1~V. Some designs propose one IVR in every die layer of the package, whereas \cite{b66} argues that a single layer dedicated to regulation would lower the resistance and free more die area. If the FIVR is placed in the substrate of the package, the motherboard and BGA connections and their parasitics are reduced, the SoC can control the regulator, local voltage domains become possible, and the distance to the load shrinks even further \cite{b60,b78}. No standard yet exists for the easy integration of the passive elements, however, and the SoC designers have to include the control system in their own design \cite{b60}. A further difficulty is that these modules must reach a current density above 1~A/mm$^{2}$ at a low height \cite{b157}.

\begin{figure}[!t]
\centerline{\includegraphics[width=\columnwidth]{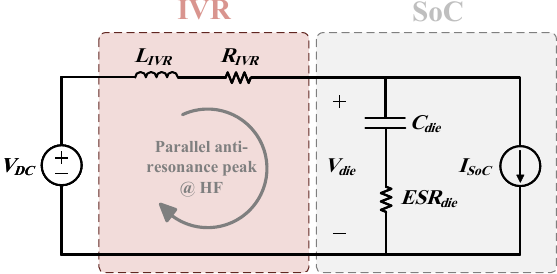}}
\caption{Equivalent circuit of the typical PDN of an on-die IVR, showing the single resonance peak at high frequency.\label{fig:16}}
\end{figure}

\begin{figure}[!t]
\centering
\panel{\columnwidth}{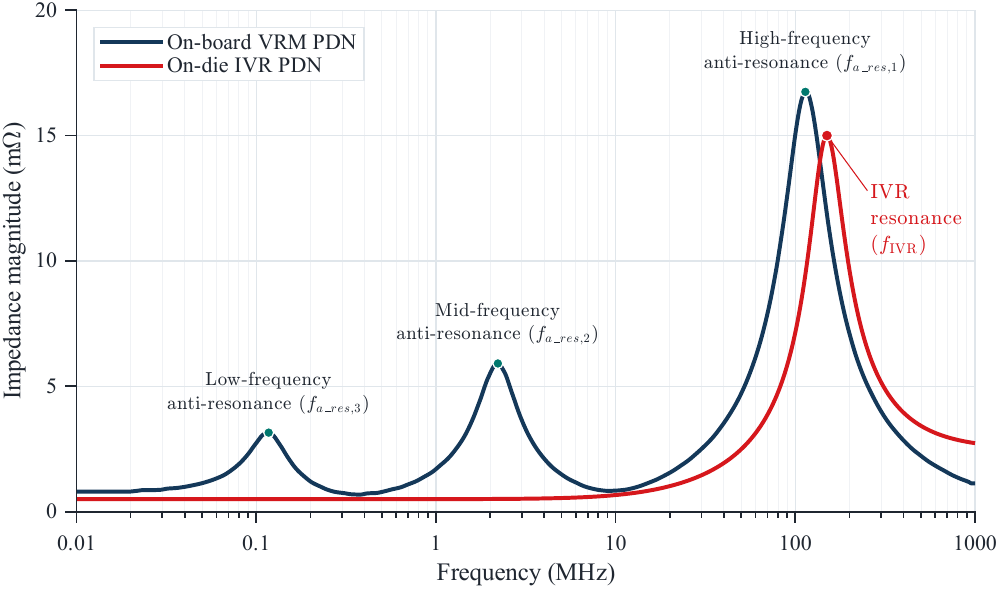}{a}\\[4pt]
\panel{\columnwidth}{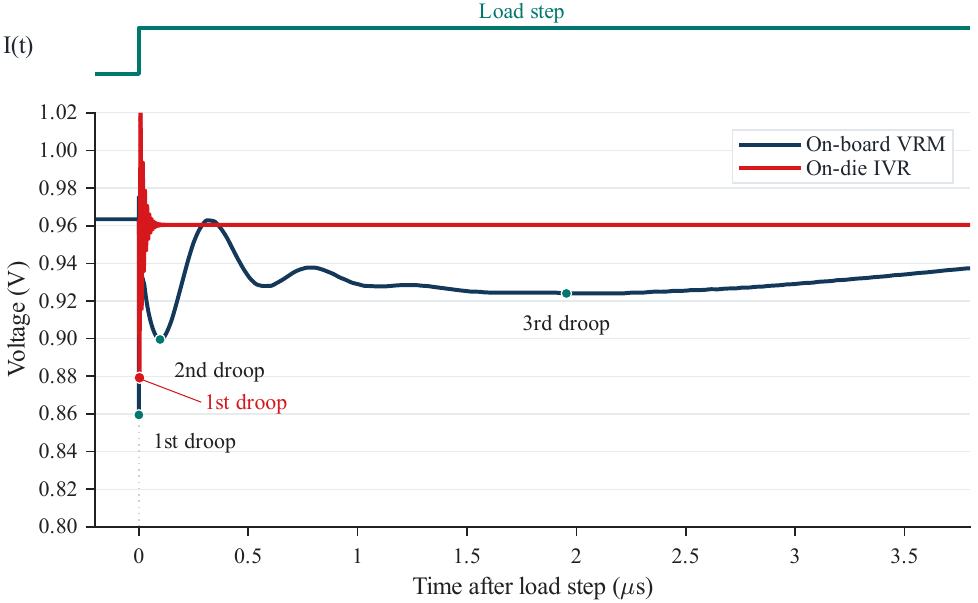}{b}
\caption{Effect of regulator integration on the PDN: comparison between a PDN with an on-board VRM and a PDN with an on-die IVR. (a) Impedance magnitude versus frequency. (b) Die voltage after a load step.\label{fig:17}}
\end{figure}

The clearest gain appears in the PDN. When the FIVR is placed on the die of the SoC, the path from the regulator to the load becomes very short, and its resistance falls below that of an on-package or on-interposer regulator \cite{b58,b158,b159}. The short path raises the density \cite{b77}, allows a higher input voltage \cite{b150}, and bypasses the socket and package parasitics of the lateral path, which are the elements that create the low- and medium-frequency anti-resonances \cite{b58,b159}. The PDN of an on-die IVR, shown in Fig.~\ref{fig:16}, is therefore left with a single anti-resonant frequency,
\begin{equation}
f_{\mathrm{IVR}}=\frac{1}{2\pi\sqrt{L_{\mathrm{IVR}}\,C_{\mathrm{die}}}}
\label{eq:ffivr}
\end{equation}
where $L_{\mathrm{IVR}}$ and $C_{\mathrm{die}}$ are the inductance of the IVR and the capacitance of the die, respectively. With a single peak, the rogue wave effect of a conventional PDN can no longer occur \cite{b33,b58,b159}. The difference between an on-board and an on-die regulator is compared in Fig.~\ref{fig:17} in both the frequency domain and the time domain.

The closest a FIVR can be brought to the die is by embedding it, and several embedding options exist \cite{b68}.
\begin{enumerate}
\item \textit{Inside the package core}: The height of the regulator is limited by the thickness of the core \cite{b68}.
\item \textit{Through the BEOL}: Cheaper than an interposer, but with fewer electrical connections \cite{b68}.
\item \textit{In an active interposer}: More expensive than the BEOL, but with more electrical connections available \cite{b68}.
\item \textit{Between the copper bumps}: Implemented with either single-sided or double-sided terminals \cite{b68}.
\end{enumerate}

Whichever option is chosen, embedding removes the need for backside cooling \cite{b104,b131}, and it brings density without soldering, lower parasitics, stacking options, the integration of active and passive components, and higher electrical and mechanical reliability \cite{b65}. To make the integration possible, the components have to be very small and dense and located close to the load \cite{b33}, and the large output filter components of a conventional VRM or IVR are far too big for this purpose \cite{b20}. An embedded FIVR can still suffer PDN loss when its core is kept relatively thick to preserve the structural integrity of the package and prevent warpage \cite{b68,b97}, and this loss accumulates along the whole path from the C4 bumps at the bottom to the micro bumps that connect the die \cite{b68}. Although the goal is to use the smallest possible components, the micro bumps and the inductors must still be able to carry the current asked of them \cite{b20}. The passive components are therefore the next thing that has to shrink.

\subsection{Embedded Inductors}
The inductor that goes into the package or onto the die is the most difficult part of an IVR design \cite{b20}, and inductors are the first candidates for embedding because they take up the largest share of the space \cite{b127,b128,b146}. They are asked to meet strict demands for high inductance density, high current density, high switching frequency, and low DC resistance (DCR) \cite{b160}. Representative targets for a 48~V to 1~V IVR are an inductance of 7--70~nH at 10--100~MHz, an inductance density above 10~nH/mm$^{2}$, a current density above 2~A/mm$^{2}$, and a DCR below 20~m$\Omega$ \cite{b160}. The inductors also have to be kept away from the high-speed I/O signals, which are sensitive to noise \cite{b20}. Larger magnetic components perform better \cite{b161} and carry more power \cite{b162}, so the package and die inductors give up part of that potential in exchange for size. The most effective way to shrink the passive components is to raise the switching frequency \cite{b51,b76,b83}, and IVRs switch at 10~MHz and above for this reason \cite{b20}. A higher frequency helps the low-energy passives, but it also raises the switching loss and lowers the current that can be delivered, which can affect the efficiency \cite{b114,b157,b163}. In return, the bandwidth of the regulator rises, which can reduce the number of output capacitors needed \cite{b53}.

The inductor may use an air core (ACI) or a magnetic core \cite{b20,b164}. An ACI can be placed on the die, but it is unlikely to reach the power conversion density required by today's IVRs \cite{b20,b160}. If the switching frequency is high enough, it can instead be built into the package from traces and vias, but the frequency it requires is above 50~MHz, which makes high-voltage conversion difficult \cite{b20}. Embedded ACIs are limited to about 10~nH and a peak current of about 20~A \cite{b79,b86,b133}, and because nothing constrains the magnetic field in the absence of a core, their scalability is limited by electromagnetic interference (EMI), radio-frequency interference (RFI), and noise coupling \cite{b20}. For a high-ratio conversion such as 48~V to 1~V, the inductance of an air core is simply too low, and a magnetic core becomes necessary \cite{b160}. Since the ACI is not always the optimal choice, attention has turned to magnetic cores, which reduce the space taken by the IVR and the distance between the regulator and the high-current point of load (PoL) \cite{b84}.

\begin{figure*}[!t]
\centering
\panel{\textwidth}{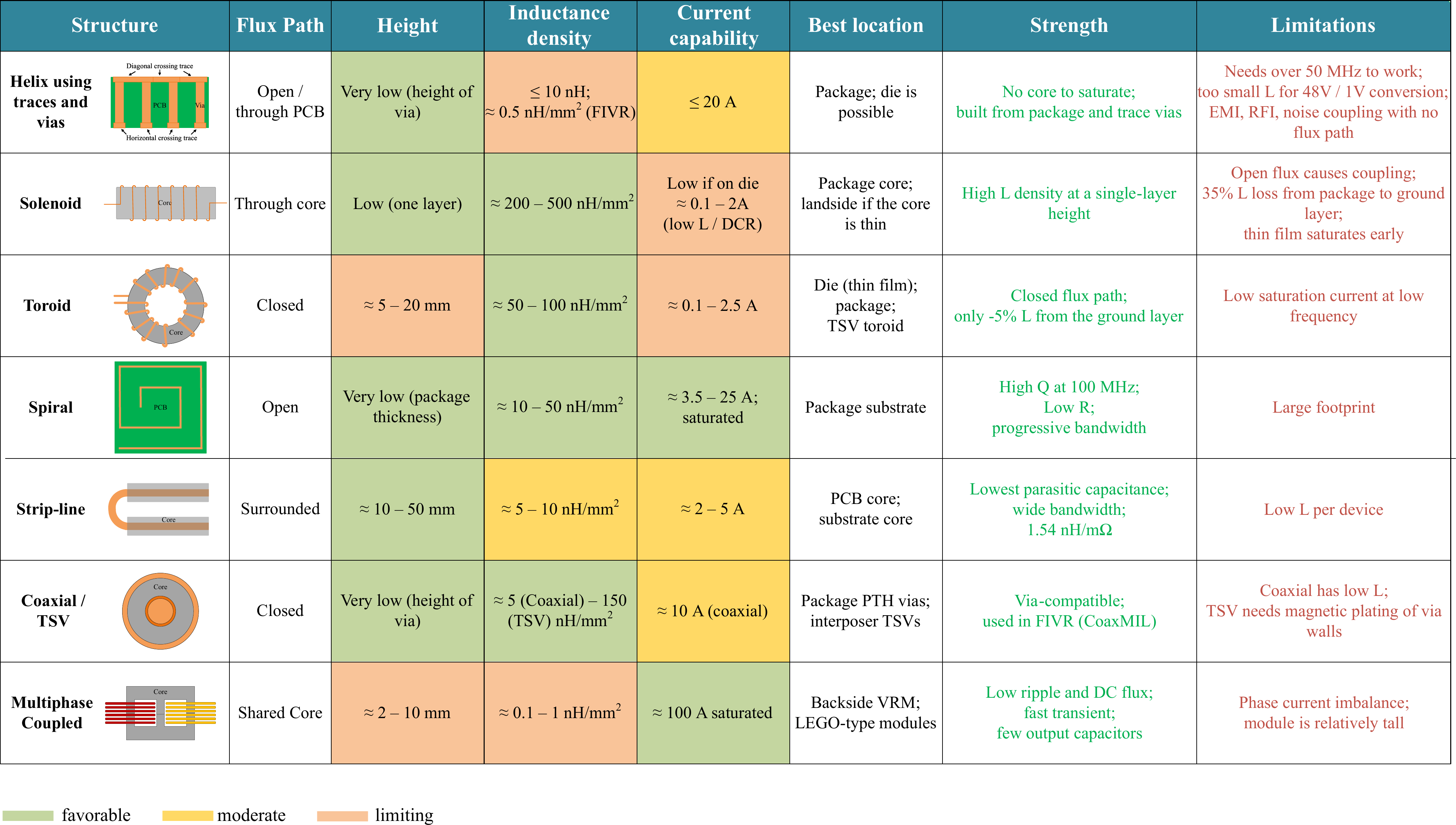}{a}\\[4pt]
\panel{\textwidth}{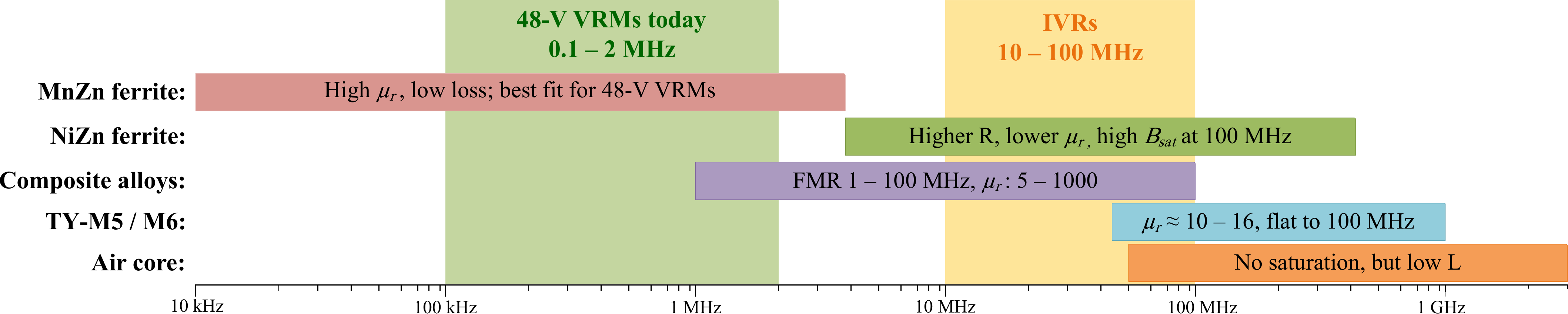}{b}
\caption{Selection map for embedded IVR inductors: (a) structures, (b) core materials.\label{fig:18}}
\end{figure*}

The choice of core material depends on the intended use of the inductor \cite{b84}, and the deciding factors are the operating frequency range, the maximum and minimum peak flux densities, the saturation flux density of the material, and its relative permeability \cite{b165}. The available materials include ferromagnetic metal alloys and nanogranular thin films, laminated and nanogranular multilayers, ferrite thin films, tapes, composites, and iron powders \cite{b157,b166}, with Intel's iron-alloy-based magnetic microparticle material, Intel's composite magnetic material, and Panasonic's HBS1 as specific examples \cite{b167}. Their relative permeability ranges from about 5 to 1000, their magnetic loss tangent from 0.01 to 1, and their ferromagnetic resonance (FMR) frequency from 1~MHz to 100~MHz \cite{b20}. An iron powder core can reach 1.6~A/mm$^{2}$ at a height as low as 5~mm, although its flux distribution lowers the inductance as the current changes \cite{b151}. Among the ferrites, manganese-zinc (MnZn) and nickel-zinc (NiZn) are the two leading options for 48~V VRMs \cite{b84}. MnZn is useful from 10~kHz to 5~MHz, with a high relative permeability and a low power loss \cite{b168,b169}, and it is the best fit for the 48~V VRMs of today, which operate between about 100~kHz and 2~MHz \cite{b84}. NiZn covers the higher range from 5~MHz to 500~MHz, with a higher resistivity and a lower relative permeability than MnZn \cite{b170}, and it is the material chosen for 100~MHz embedded inductors because its permeability stays stable over a wide range of field intensities and frequencies, its saturation point is high, and its core loss is low \cite{b160}. Pre-release materials developed specifically for 100~MHz IVRs show a relative permeability of about 10 to 16 that stays flat up to several hundred megahertz, an FMR near 2~GHz, and a permeability that drops to 70\% of its initial value only at flux densities of about 0.7--0.8~T \cite{b167}. High-permeability materials, on the other hand, have a loss tangent at 10~MHz and above that makes them impractical for very high frequencies \cite{b20}. For multiphase coupled inductors, cores of cobalt-zirconium-tantalum (CoZrTa) or Permalloy help to raise the flux, and lamination limits the eddy currents that highly permeable materials would otherwise produce \cite{b20,b157}. Saturation can be avoided by air gaps between the two magnetic layers \cite{b20,b81,b171} or by specific winding configurations \cite{b163}.

Inductors may be coupled or uncoupled. At low currents, an uncoupled inductor can develop a DC flux that saturates its core \cite{b20}. A coupled inductor reduces the current ripple of an uncoupled one, which allows smaller output capacitors, and it also reduces the leakage inductance and gives a fast initial response to a current demand \cite{b81,b124,b163}. The reduction in output capacitance is also helped by the parallel connection of the inductors, which lowers the impedance \cite{b53}. For small, high-performance IVRs, multiphase coupled inductors are the suitable choice, since they reduce the DC flux in the magnetic core and improve the transient response \cite{b113,b126,b172,b173,b174}. The faster transient allows a higher control bandwidth, which reduces the number of decoupling capacitors, and it lowers the voltage droop, so that the designer can reduce the voltage margin kept inside the 5--10\% band \cite{b53}. The combination of a higher switching frequency and a magnetic core produces inductors of about 1--2~nH \cite{b20,b79}, and in the case of Ponte Vecchio the values reported are currents up to 8~A, an inductance of at most 900~nH, and a resistance down to 12~m$\Omega$ \cite{b79,b148}. Current-sharing arrangements of this type are not immune to phase current imbalance, which can appear when they resonate in some circuits \cite{b106}. The trans-inductor voltage regulator (TLVR) uses multiphase coupling so that the coupling inductance differs between the steady state and the switching state \cite{b106}, and the Linearly Extendable Group Oriented (LEGO) converter of \cite{b162} uses a four-phase vertical coupled inductor in which the core is wrapped around the winding instead of the winding around the core \cite{b162,b12}.

\begin{table*}[!t]
\caption{Embedded Inductors Reported for Integrated Voltage Regulators}
\label{tab:ind}
\setlength{\tabcolsep}{2pt}
\renewcommand{\arraystretch}{1.25}
\begin{tabular*}{\textwidth}{@{\extracolsep{\fill}}cclcccccccc}
\hline\noalign{\vskip 2pt}
\tabhead{Ref.} & \tabhead{Year} & \tabhead{Inductor structure} & \tabhead{Core material\\($\mu_{r}$)} & \tabhead{$f_{\mathrm{sw}}$\\(MHz)} & \tabhead{$L$\\(nH)} & \tabhead{DCR\\(m$\Omega$)} & \tabhead{$I_{\mathrm{sat}}$\\(A)} & \tabhead{Footprint\\(mm$^{2}$)} & \tabhead{$L$/Area\\(nH/mm$^{2}$)} & \tabhead{$L$/DCR\\(nH/m$\Omega$)} \\
\noalign{\vskip 2pt}\hline
\cite{b34} & 2014 & Air-core FIVR inductor & Air, 1 & 140 & 1.2 & 7 & 8.0 & 2.0 & 0.6 & 0.17 \\
\cite{b176} & 2021 & Package-embedded toroid & Composite, 180 & 1 & 480 & 23 & 0.1 & 5.0 & 96 & 20.9 \\
\cite{b160} & 2025 & Progressively widened spiral, in package & NiZn, 4.7 & 100 & 78 & 11 & 24 & 7.02 & 11.1 & 7.1 \\
\cite{b167} & 2025 & Strip-line in PCB or substrate core & TY-M5, $\approx$16 & 100 & 9.84 & 6.4 & -- & 1.0 & 9.84 & 1.54 \\
\cite{b167} & 2025 & Strip-line in PCB or substrate core & TY-M6, $\approx$10 & 100 & 7.75 & 6.8 & -- & 1.0 & 7.75 & 1.14 \\
\cite{b177} & 2022 & Package-embedded toroid & 30 & 1 & 420 & 89 & 2.5 & 8.8 & 48 & 4.7 \\
\cite{b146} & 2025 & Dual-core single spiral & 90 & 2 & 115 & 30 & 5.0 & 8.8 & 13 & 3.8 \\
\cite{b146} & 2025 & Two parallel spirals & 90 & 2 & 290 & 85 & 3.5 & 9.1 & 32 & 3.4 \\
\cite{b146} & 2025 & Two parallel spirals & 180 & 2 & 330 & 85 & 3.5 & 8.9 & 37 & 3.9 \\
\cite{b178} & 2020 & Strip-line magnetic inductor array & Fe-alloy microparticle & 100 & 3.0 & 12 & 4.0 & 0.5 & 6.0 & 0.25 \\
\cite{b179} & 2021 & Coaxial magnetic composite core & Composite, 8.5 & 90--100 & 2.5 & 12 & 8.0 & 0.4 & 6.3 & 0.21 \\
\cite{b180} & 2013 & Coupled magnetic core on silicon interposer & -- & 100 & 5.0 & 270 & -- & 0.245 & 20.4 & 0.019 \\
\hline
\end{tabular*}
\end{table*}

The flux may travel through the core laterally or vertically \cite{b162}. A horizontal structure runs the flux horizontally through the windings, whereas a vertical structure has both horizontal and vertical paths as the current travels through the wires \cite{b162}. The vertical windings are short, and their reduced path lowers the parasitics and improves the transient performance \cite{b124}. Reference~\cite{b127} argues that the vertical flux structure has more advantages than the horizontal one, and its inductor reached a saturation current of 110~A and a peak current density of 1.9~A/mm$^{2}$ at a height of only 2.2~mm \cite{b127}.

IVR inductors are built in several shapes, namely the solenoid, the toroid, the spiral, the multiphase coupled, the strip-line, and the coaxial structure \cite{b167,b175}. Their properties are summarized in Fig.~\ref{fig:18} and Table~\ref{tab:ind}, and the main ones are the following.

\begin{enumerate}
\item \textit{Solenoid}: A high inductance density is obtained with a single layer, which gives a very low height \cite{b20}. The flux path through the core is not closed, however, so unwanted coupling can occur, and on the die the inductor is so small that its low $L$/DCR ratio prevents a high current density \cite{b20}. The thin-film magnetic material saturates at low current, so the inductor is better placed on the core of the package, where the thick core raises the current capability; if the core is absent or ultra-thin, the landside of the package is the better location \cite{b20}. The ground layer of the package reduces the inductance of a solenoid by about 35\% \cite{b20}.
\item \textit{Toroid}: The flux path is closed and the structure is short \cite{b146}, which makes it a good candidate for IVRs \cite{b157}, and it is the common choice for thin-film inductors on the die \cite{b20}. The ground layer reduces its inductance by only about 5\% \cite{b20}. A toroidal inductor can also be built from TSVs in a 3-D chip or in the redistribution layer of a 2.5-D chip, and adding a magnetic core with air gaps improves it further \cite{b78}.
\item \textit{Spiral}: An overlapped dual-core spiral uses the mutual inductance between its cores to raise the inductance density to 37~nH/mm$^{2}$ \cite{b146}. A spiral with progressively widened turns and an aspect ratio between 1 and 10, which exploits the thickness of the package, reaches a quality factor of about 105 at 100~MHz with 78~nH, 11~m$\Omega$, a saturation current of 24~A, and a footprint of about 7~mm$^{2}$, and it delivers 10~A with an inductor efficiency of about 90\% \cite{b154}.
\item \textit{Strip-line}: A single strip surrounded by magnetic material has a much lower parasitic capacitance than a multiturn structure, which gives it a wider bandwidth \cite{b161}. With a height of 0.35~mm it can be embedded in the core of the PCB or of the package substrate, and with the new 100~MHz materials it reaches 9.84~nH at 6.4~m$\Omega$, or 1.54~nH/m$\Omega$, with a calculated inductor efficiency of 97.5\% when converting 1.8~V to 0.9~V at 100~MHz and 2~A \cite{b161}.
\item \textit{Coaxial and via-based}: Coaxial magnetic integrated inductors (CoaxMILs) built with PTH vias have allowed FIVRs to be implemented \cite{b107,b104}. A TSV inductor whose via walls are coated with electrodeposited magnetic material reaches 132~nH/mm$^{2}$ \cite{b83}. A customer-specific embeddable inductor with via-based connections covers 5--100~MHz with 20~nH/mm$^{2}$ and a self-resonant frequency (SRF) at or above 10~MHz \cite{b77}, and the silicon-based ultimate miniature magnetic inductors and transformers (SUMMIT) of \cite{b35} are far-BEOL devices only 75~$\mu$m high, with 25~$\mu$m of copper and a magnetic core that give low resistance and high inductance \cite{b35}. Such devices can be embedded with the existing substrate manufacturing processes \cite{b146}.
\item \textit{Matrix planar transformer}: For the LLC DC transformer (DCX) used in HPC, the matrix planar transformer lowers the AC and leakage-inductance losses and lets the rectifying switches and the filter capacitors sit inside the secondary windings \cite{b28}, and fractional-turn transformers have been studied for the same reason \cite{b28}. A two-dimensionally scalable design with cross-shaped cores and the output terminals integrated inside the transformer, built for mounting on the back of the processor, delivers 450~A and 810~W from 58~V to 1.8~V with a peak efficiency of 94.4\%, a transformer power density of 1660~W/in$^{3}$, and a converter power density of 924~W/in$^{3}$ at a height of 8~mm and 0.25~A/mm$^{2}$ \cite{b28}; the design also appears in Table~\ref{tab:vpd}.
\end{enumerate}
Across all of these structures, the embedded inductors reported so far handle current densities of up to about 1~A/mm$^{2}$, and improvements in inductor and capacitor technology are expected to raise this limit \cite{b30}. In every case, the way in which heat is generated and removed has to be kept in mind so that nothing overheats \cite{b20}.

\subsection{Embedded Capacitors and Switches}
The switches of an IVR are the least of its problems. Power field-effect transistors (FETs) in CMOS operate comfortably above 10~MHz \cite{b20}. Their voltage rating, however, is only about 1~V, so a FIVR with a 1.8~V input has to use cascoded switches \cite{b20,b34}. Stacking switches raises the input voltage and lowers the current, but it costs more switches, more on-resistance, and more area \cite{b20}. Gallium nitride (GaN) switches, which offer a higher bandwidth than silicon, bring the possibility of controlling higher voltages at higher temperatures \cite{b20}, and as IVRs move to higher voltages, FETs and GaN switches are targeting switching frequencies of 5--50~MHz, above which the switching loss becomes excessive \cite{b20}. GaN has the best switching performance, silicon carbide (SiC) performs best at high temperature, and silicon (Si) has the lowest cost, which is why all three are used in the 3.3--12~kW rack PSUs built for AI \cite{b181}. Combining GaN switches with topologies that shrink the passive components and with better magnetic materials is what will allow the IVR to become denser and more efficient \cite{b84}.

\begin{figure*}[!t]
\centering
\panel{\textwidth}{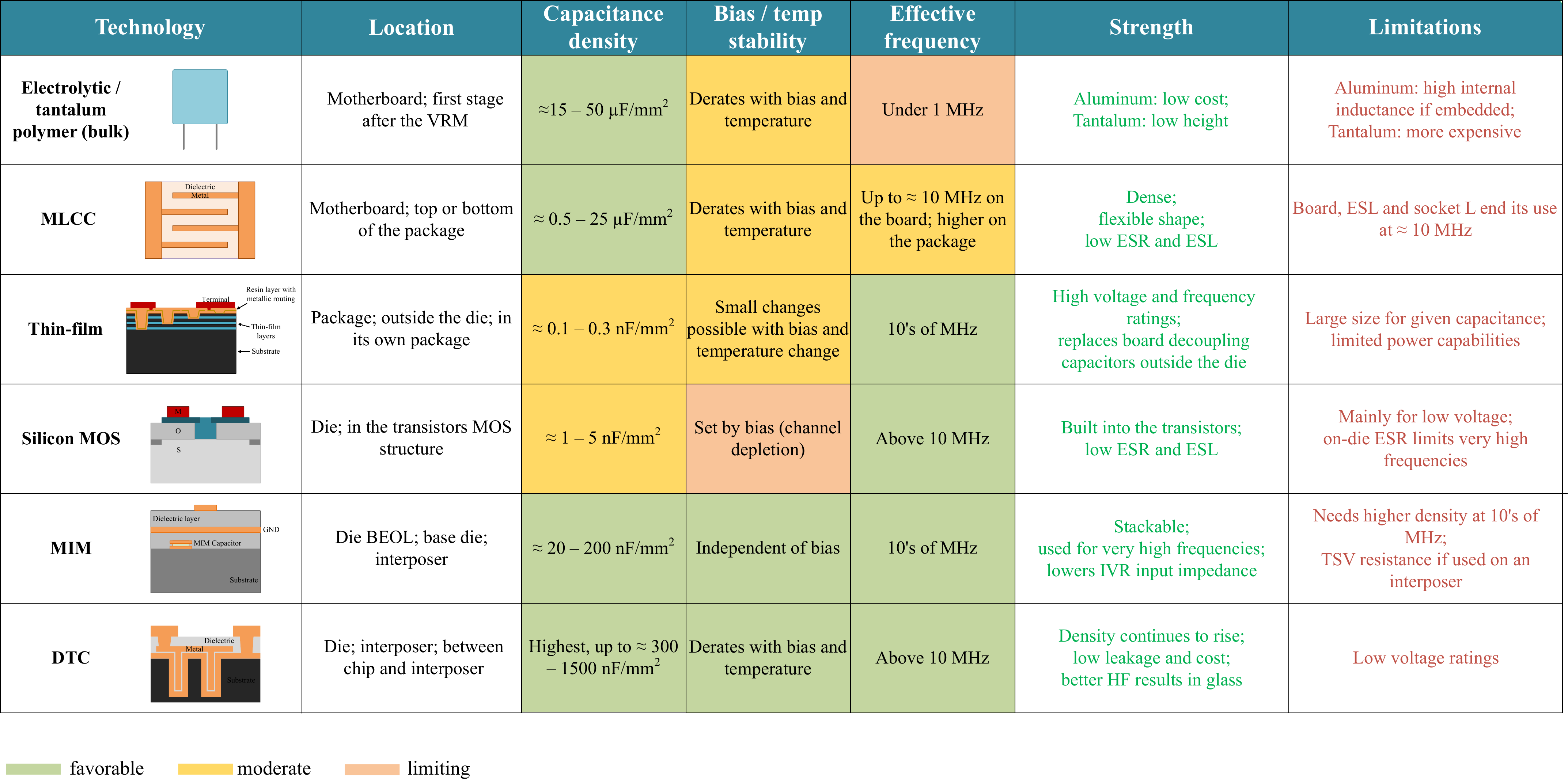}{a}\\[4pt]
\panel{\textwidth}{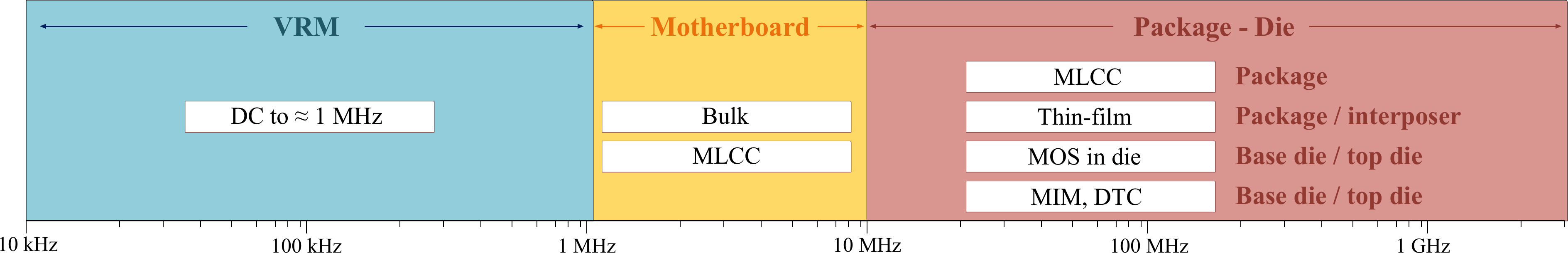}{b}
\caption{Selection map for decoupling and embedded capacitors: (a) technologies, (b) placement.\label{fig:19}}
\end{figure*}

Bringing the PDN closer to the load to reduce the loop inductance also opens the way to embedded capacitors for output filtering, in place of the discrete capacitors of the motherboard \cite{b33,b20}. Embedded capacitors give better electrical performance, a lower height, and a shorter PDN path \cite{b59,b182}. Not all of them qualify, since embedded film capacitors are not very compact and embedded aluminum electrolytic capacitors have a high internal inductance and poor high-frequency performance \cite{b59}. The candidates that remain are compared in Fig.~\ref{fig:19}.

\begin{enumerate}
\item \textit{Thin-film and embedded array capacitors}: Outside the die, they are suitable replacements for the silicon board decoupling capacitors because of their high voltage and frequency ratings, and they come in their own package \cite{b20}.
\item \textit{MIM capacitors}: Their capacitance density is higher than that of MOS capacitors, and they can be stacked in the back-end metal layers of the die to raise it further \cite{b59,b66,b183}. They act as a decoupler that lowers the input impedance of the IVR \cite{b33}, their capacitance does not change with the bias voltage as that of a MOS capacitor does, and in the time domain their effect is seen as a reduction of the first and largest voltage droop \cite{b20}. They operate at high frequency, but with switching frequencies in the tens of megahertz they need a still higher density to hold the voltage within its band \cite{b20}. They also form the flying capacitors of the monolithic SCVRs described in Section~\ref{sec:ivr}-\ref{sec:ivr-levels} \cite{b150}.
\item \textit{Deep trench capacitors (DTCs)}: Their capacitance density exceeds that of the MIM capacitor \cite{b20,b59}. They can be placed on the die, on the interposer, or made so short that they fit between the chip and the interposer in place of some of the BGA solder balls \cite{b57,b102}. They filter from 10~MHz upward, and their density keeps rising \cite{b138}; a high-k (HK) DTC with a capacitance density of up to 340~nF/mm$^{2}$, low leakage, and low cost has been manufactured for operation at 1.35~V \cite{b57}. Embedded in glass, the DTC has also shown improved high-frequency performance in radio-frequency (RF) applications \cite{b59}.
\end{enumerate}

MIM and deep trench (DT) capacitors could together replace the MLCCs for all mid- and high-frequency decoupling \cite{b20}. Reference~\cite{b104} connects fin field-effect transistors (FinFETs) to MIM capacitors through three metal layers in a 12~nm node to improve current density and efficiency. The growing density of the MIM capacitor has also made a new topology possible, the continuous capacitive voltage regulator (C2VR), which regulates through a continuously scalable conversion ratio (CSCR) block placed between two SC circuits \cite{b107,b104}. A silicon interposer allows MIM and DT capacitors to be used with the IVR, which is favorable for the signals, but from the power integrity side it requires TSVs, whose resistance is higher than that of a copper micro bump and appears as loss in the PDN \cite{b20}. Using embedded inductors and capacitors together allows a higher input voltage, with high-frequency metal--oxide--semiconductor FETs (MOSFETs) or low-voltage GaN switches and a multilayer core in which the passives are stacked \cite{b104}. Foveros allows the IVR to sit directly under the load on the top die while the MIM and DT capacitors are integrated into the base die \cite{b20}, and instead of patterning many layers of electrodes inside the wafer, the capacitors can be embedded simply by etching a cavity for them \cite{b128}. Reference~\cite{b77} has also introduced a customer-specific embeddable capacitor rated at 2~V, with a capacitance density above 4~$\mu$F/mm$^{2}$, an ESL below 50~pH, a series resonant frequency of 1--5~MHz, a wide attenuation band, and no temperature or bias derating \cite{b77}.

\subsection{Thermal Management of Embedded Regulators}
Smaller passive components come at a thermal price \cite{b33,b68}. An embedded inductor needs a magnetic core to beat the density of an air core, but the core adds Joule heating and a temperature limit, since a ferrite core loses its magnetic properties beyond its Curie temperature \cite{b33}. An inductor embedded in the PCB also sits closer to other heat sources such as the switches, and the thermal coupling raises its temperature further \cite{b20,b33}. Air cooling with fans handles power densities of up to about 1.5~W/mm$^{2}$ and microchannel cooling up to about 7.9~W/mm$^{2}$ \cite{b33,b184}, whereas a FIVR integrated on the die, where the switches and the control are tightly grouped, exceeds 10~W/mm$^{2}$, well above conventional VPD \cite{b33}. Liquid cooling and liquid immersion cooling are possible in AI data centers despite their cost \cite{b20}, and immersion cooling is now becoming standard \cite{b64}. A two-stage IVR lets the intermediate bus converter (IBC) share the immersion cooling whether the second stage sits on the bottom of the motherboard or between the motherboard and the package \cite{b43}. An embedded inductor, however, can become thermally insulated, because neither air nor immersion fluid can reach through the BGA isolation \cite{b46}, and a specialized cooling solution is then required \cite{b33}. If the components are not cooled, the efficiency falls and the converter may fail \cite{b46}. Keeping the regulator on the backside of the motherboard avoids this isolation altogether \cite{b46}. Heat sinks, heat pipes, and vapor chambers are used to keep the die junction below its maximum temperature \cite{b128}, and if the temperature feedback shows the junction above the usual limit of 100~$^{\circ}$C, the DVFS lowers the frequency and the voltage to cool the die \cite{b20}. On a silicon interposer, where thermal coupling is a concern, thermal metamaterials, an air cavity, or a glass substrate can help to spread the heat \cite{b33,b185,b186}.

\subsection{Commercial Solutions}
Several of these ideas have already reached products. The PowerTile of \cite{b187} is an integrated solution that delivers 1000~A and can be scaled beyond 10\,000~A. The Crescendo line of \cite{b188} scales to 2600~A and includes the EP70xx and EP71xx IVR series together with embedded capacitors based on its FinFast technology. Reference~\cite{b189} reports a four-phase FIVR with a current density of 2~A/mm$^{2}$ in its TDM2454xx modules, and its Fe1736 module is a 16-phase buck converter that converts 1.8~V to 1.5--0.25~V with a peak current of 56~A at a height of only 0.6~mm. Reference~\cite{b190} shows how FPA can respond in less than a microsecond with 100~A. Solutions described as 25 times smaller, more than 100 times faster, and more than 30\% more efficient than the existing ones have also been claimed \cite{b189}.

The IVR, its fully integrated form, and the embedded passives described in this section are the means by which the vertical path of Section~\ref{sec:vpd} is shortened to its limit. The gain is a PDN with a single anti-resonance and a regulator that responds within the die. The cost is an inductor that must be designed for the package, capacitors that must be built into the silicon, and heat that must be removed from a space that no coolant can reach. The next section quantifies the gain with a power integrity simulation.

\begin{figure*}[!t]
\centering
\panel{0.49\textwidth}{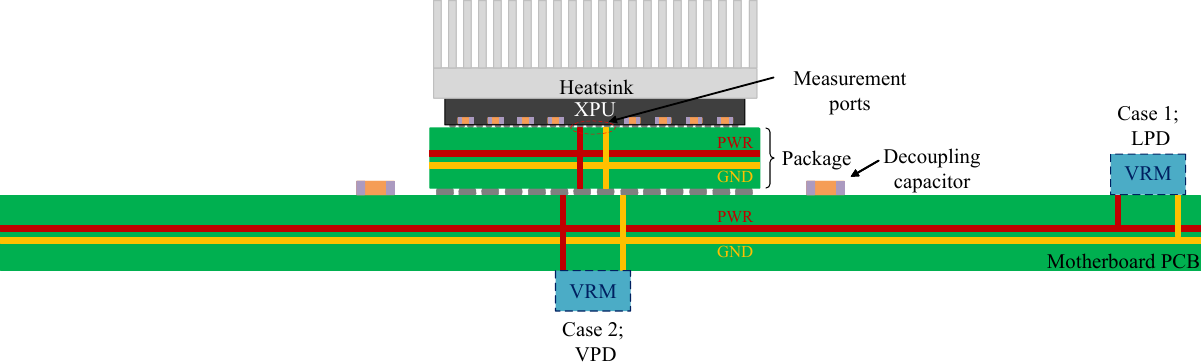}{a}\hfill\panel{0.49\textwidth}{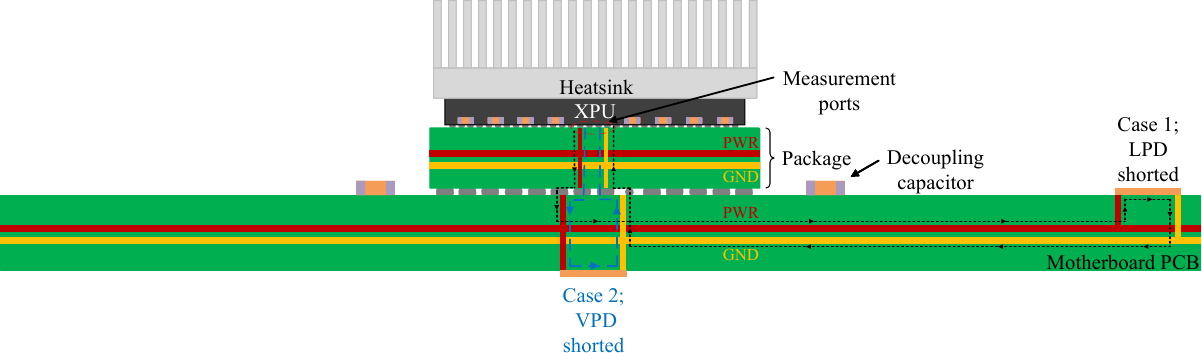}{b}\\[4pt]
\panel{0.49\textwidth}{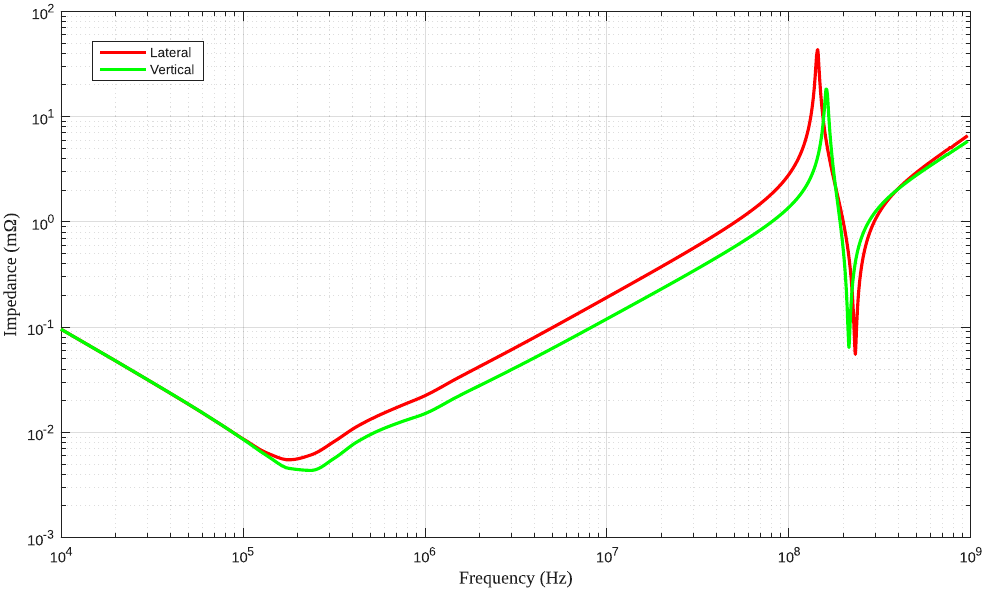}{c}\hfill\panel{0.49\textwidth}{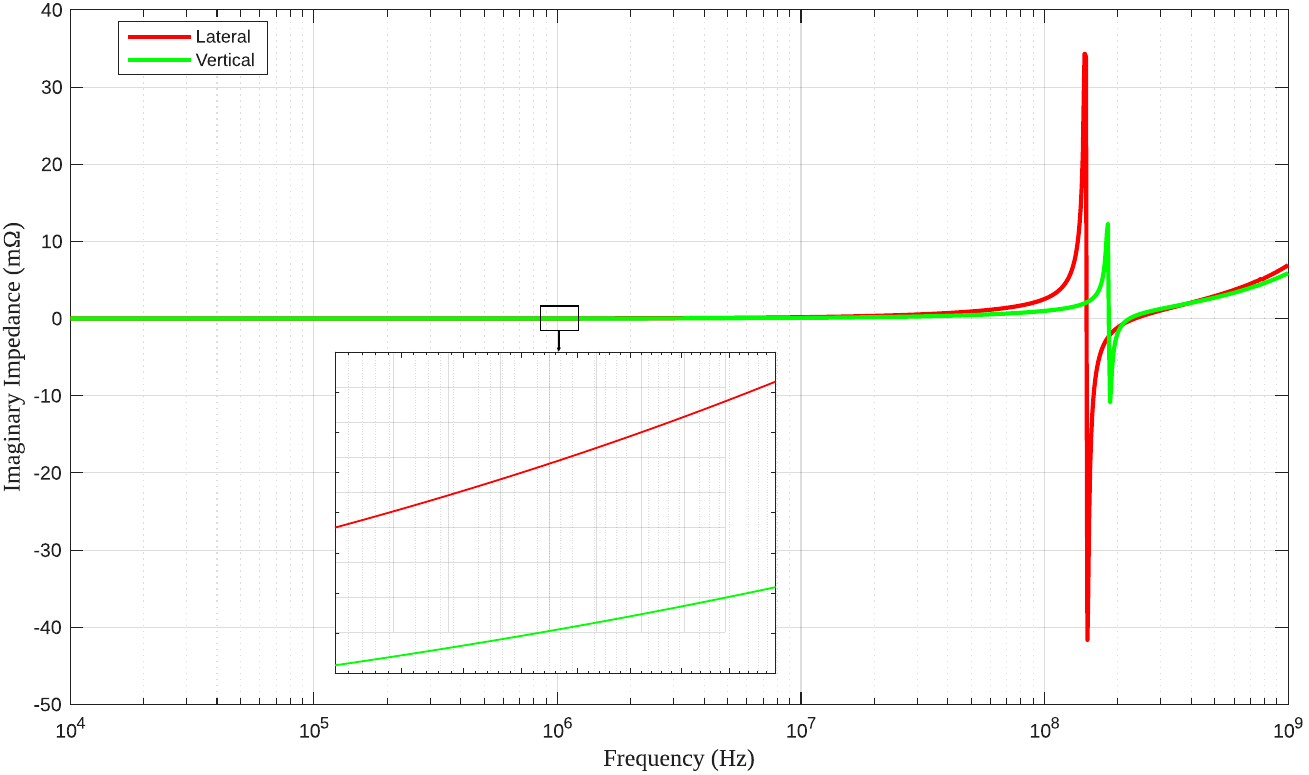}{d}
\caption{Circuit board layouts and impedance comparison for a single VRM. (a) VRM placed laterally at the far end of the board (Case 1) and vertically below the die (Case 2). (b) Current loops of the two cases with the VRM shorted. (c) Impedance seen at the die for the lateral and the vertical placement. (d) Imaginary impedance of the paths for lateral and vertical placements.\label{fig:20}}
\end{figure*}

\section{Power Integrity Simulations and VPD Design Guidelines}
\label{sec:sim}
The preceding sections have argued from reported designs that a shorter delivery path lowers the impedance of the PDN, and that VPD has better performance compared with LPD for powering GPUs in data centers. This section examines these claims and provides guidelines and design rules for a better PDN design. The simulations are carried out in Cadence Sigrity PowerSI, which models the PDN elements and the PCB layout. The measurement port is placed at the center of the motherboard, at the position where the die would be mounted, so that the impedance returned by the solver is the impedance seen by the load. Different VRM position scenarios are modeled, and each VRM is represented by its output capacitor. Each output capacitor is built from a netlist obtained from PSPICE. Every case is swept from 10~kHz to 1~GHz.

\subsection{Single VRM Placed Laterally and Vertically}
The first set of simulations compares LPD with VPD for a single VRM with a 220~$\mu$F output capacitor. In the lateral case, shown as Case 1 in Fig.~\ref{fig:20}(a), a VRM is placed at the far top of the motherboard and is connected to the power and ground planes at the center of the board. The simulation is then repeated with the same VRM placed below the die, shown as Case 2 in Fig.~\ref{fig:20}(a), so that the two PDN impedances can be compared over the frequency range of interest.

The resulting impedances are compared in Fig.~\ref{fig:20}(c). In both simulations the VRM output capacitor produces a negative slope at low frequency that continues down to the resonant frequency of the output capacitor. As can be seen, the resonant frequency of the vertical arrangement is higher than that of the lateral one, because its loop inductance is smaller. This means that the vertical arrangement has a better and broader low-frequency capacitive range compared with the lateral arrangement. In addition, the vertical arrangement has a better and lower resonant impedance than the lateral one, thanks to its lower delivery path resistance. From the first resonance, at about 200~kHz, the ESL of each output capacitor raises the impedance toward an anti-resonant peak, which is created by the capacitance and the inductance of the power and ground planes together with the ESL of the output capacitor. Over this range the vertical arrangement again shows lower inductance and impedance as the anti-resonant peak is approached. The plane capacitance of the PCB then brings the impedance down over its effective range until the planes reach their own resonant frequency. Beyond that point no capacitor with a smaller capacitance is left to compensate for the inductive effects that occur at higher frequencies, so the impedance begins to rise again.

\begin{table}[!t]
\caption{Comparison of Loop Inductances for Different Capacitor Distances}
\label{tab:loop}
\setlength{\tabcolsep}{3pt}
\renewcommand{\arraystretch}{1.25}
\begin{tabular*}{\columnwidth}{@{\extracolsep{\fill}}cccc}
\hline\noalign{\vskip 2pt}
\tabhead{Type of\\Delivery} & \tabhead{Distance from\\measurement\\ports (mm)} & \tabhead{Imaginary\\Impedance\\(m$\Omega$)} & \tabhead{Loop\\Inductance\\(nH)} \\
\noalign{\vskip 2pt}\hline
\textbf{LPD} & 50 & 18.903 & 3.36 \\
\textbf{LPD} & 37.5 & 17.929 & 3.21 \\
\textbf{LPD} & 25 & 17.217 & 3.12 \\
\textbf{LPD} & 12.5 & 16.189 & 2.94 \\
\textbf{VPD} & 0 & 10.362 & 1.77 \\
\hline
\end{tabular*}
\end{table}

\begin{figure*}[!t]
\centering
\panel{0.49\textwidth}{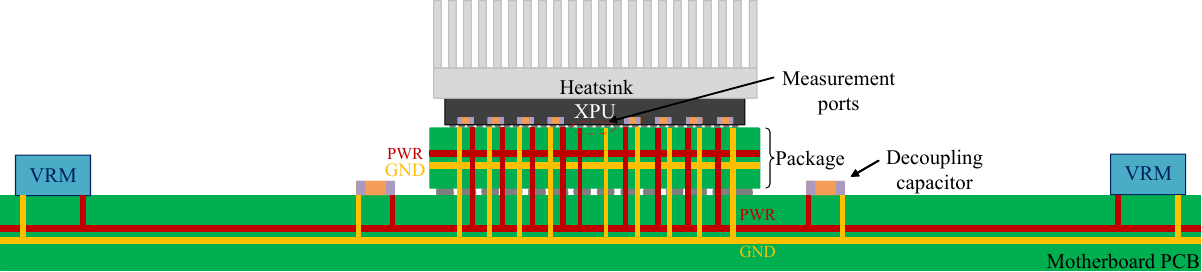}{a}\hfill\panel{0.49\textwidth}{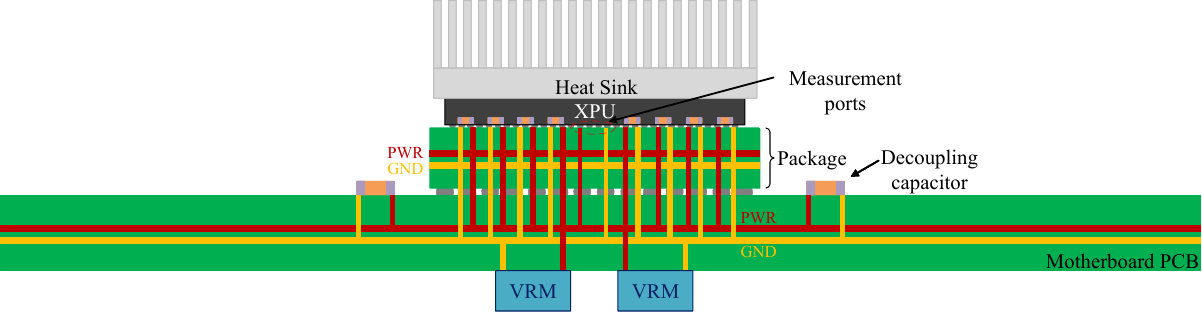}{b}\\[4pt]
\panel{0.49\textwidth}{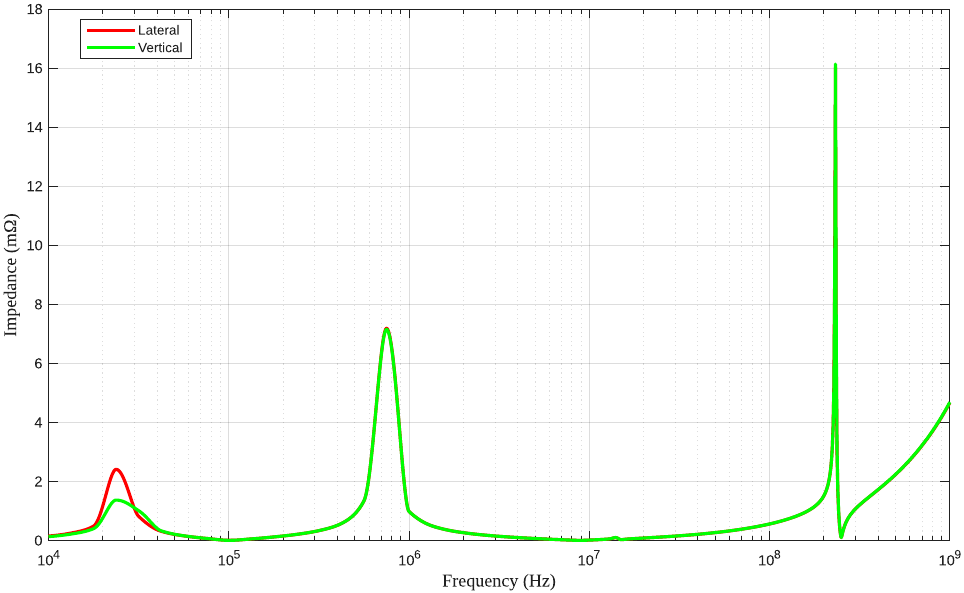}{c}\hfill\panel{0.49\textwidth}{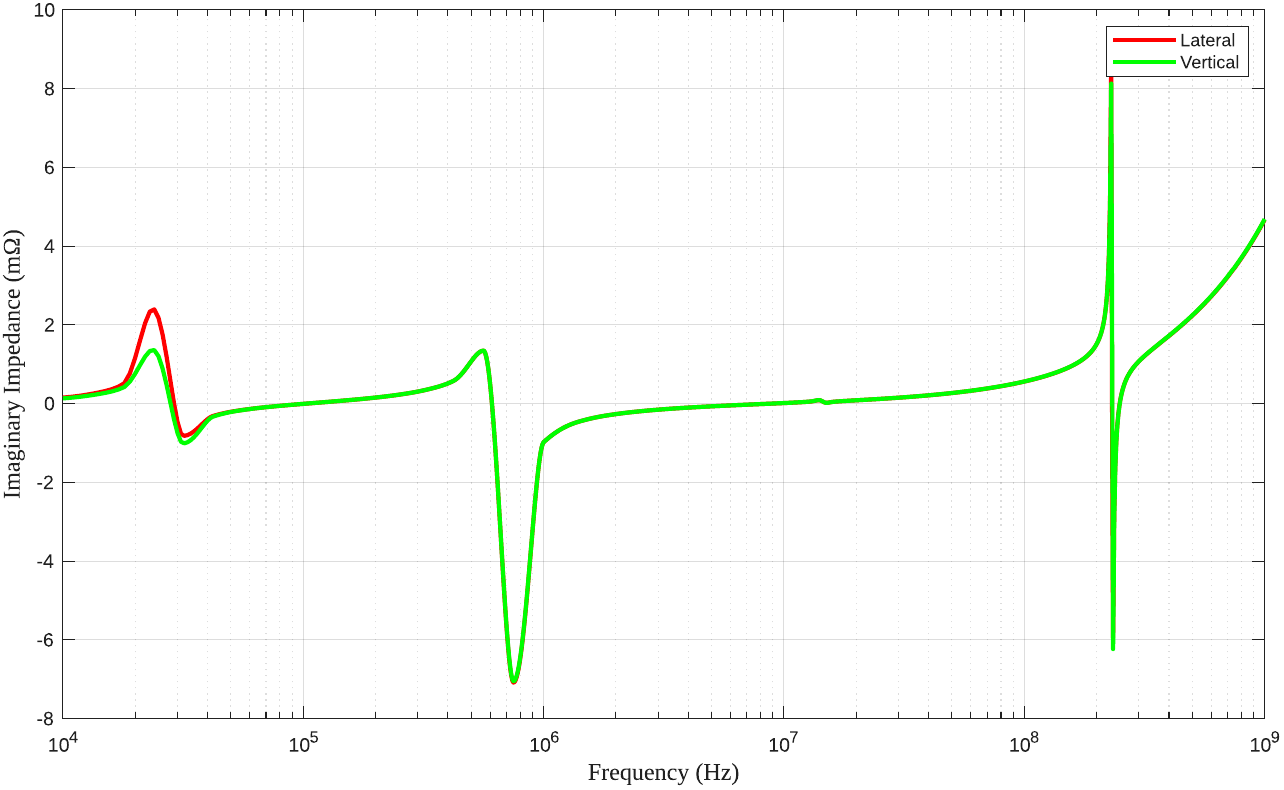}{d}
\caption{Lateral and vertical delivery in a PDN. (a) Lateral layout. (b) Vertical layout. (c) Impedance comparison. (d) Imaginary impedance comparison.\label{fig:21}}
\end{figure*}

In the next step, the loop inductance is examined along the board. Fig.~\ref{fig:20}(b) shows the current loop of each case when the VRM is shorted. A lateral VRM 50~mm away from the die is moved toward a vertical position directly underneath the measurement port in equal intervals of 12.5~mm. The loop inductance can be obtained as:
\begin{equation}
L=\frac{\left|Z_{\mathrm{im}}(f)\right|}{\omega}
\label{eq:loop}
\end{equation}
where $\left|Z_{\mathrm{im}}\right|$ is the magnitude of the imaginary impedance, $\omega$ is the radian frequency, and $L$ is the loop inductance. The imaginary impedance at 1~MHz is read for each distance, and the loop inductances obtained in this way are collected in Table~\ref{tab:loop}, which shows that the closer the VRM is to the load, the lower the loop inductance. Fig.~\ref{fig:20}(d) compares the loop inductance of the farthest lateral VRM scenario at 50~mm with that of the vertical scenario.

\subsection{Lateral and Vertical PDN}
The second set of simulations moves from a single VRM arrangement to the arrangement on which an AI motherboard is normally based. In the lateral layout shown in Fig.~\ref{fig:21}(a), the VRMs are placed laterally at the far end of the motherboard, the mid-frequency decoupling capacitors close to the SoC, and the high-frequency decoupling capacitors inside the SoC. The vertical layout, shown in Fig.~\ref{fig:21}(b), is built by relocating the VRMs from the far end of the motherboard to directly underneath the SoC. In this situation, the difference in the magnitude of the observed impedance can be attributed to that single change.

The lateral and vertical PDN impedances are compared in Fig.~\ref{fig:21}(c). Over the whole frequency range, the difference appears at the low-frequency anti-resonance, which is the expected outcome, since the VRMs are the only elements that were moved. In this range the vertical arrangement gives the lower anti-resonant peak, which is what a PDN designer seeks. This is the result of the lower path resistance of the vertical arrangement compared with the lateral one. The imaginary impedance in Fig.~\ref{fig:21}(d) confirms the same behavior. The reactive response is similar for the mid- and the high-frequency ranges and differs only at low frequency, because moving the VRMs closer to the die makes the loop inductance smaller and therefore reduces the reactive inductance.

\subsection{Different VRM Positions}
\begin{figure*}[!t]
\centering
\panel{0.49\textwidth}{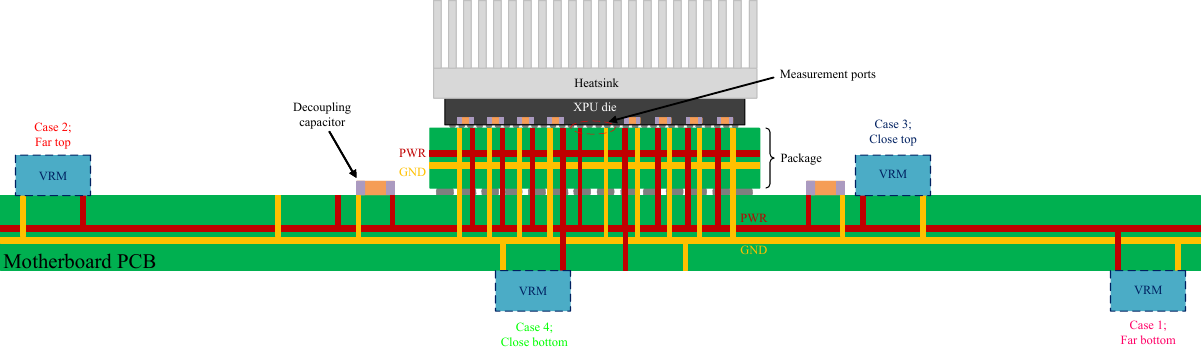}{a}\hfill\panel{0.49\textwidth}{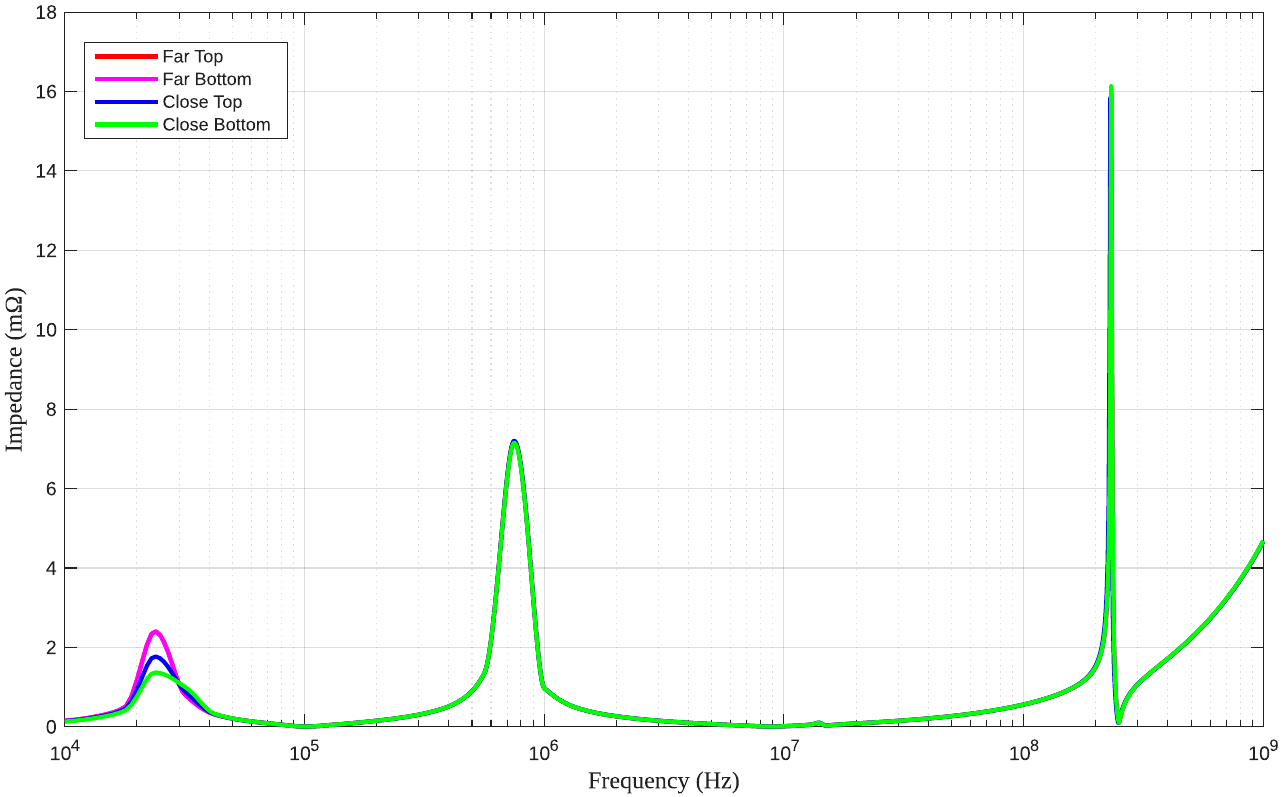}{b}\\[4pt]
\panel{0.49\textwidth}{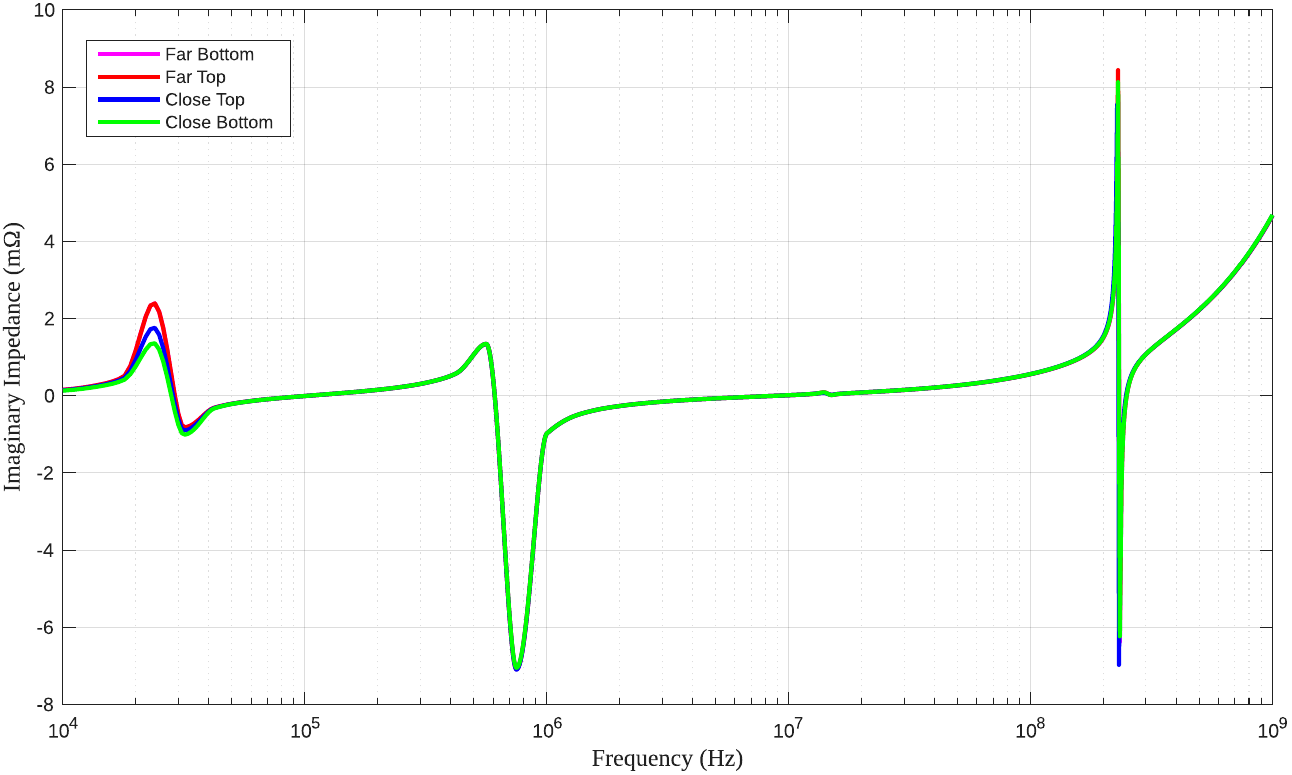}{c}\hfill\panel{0.49\textwidth}{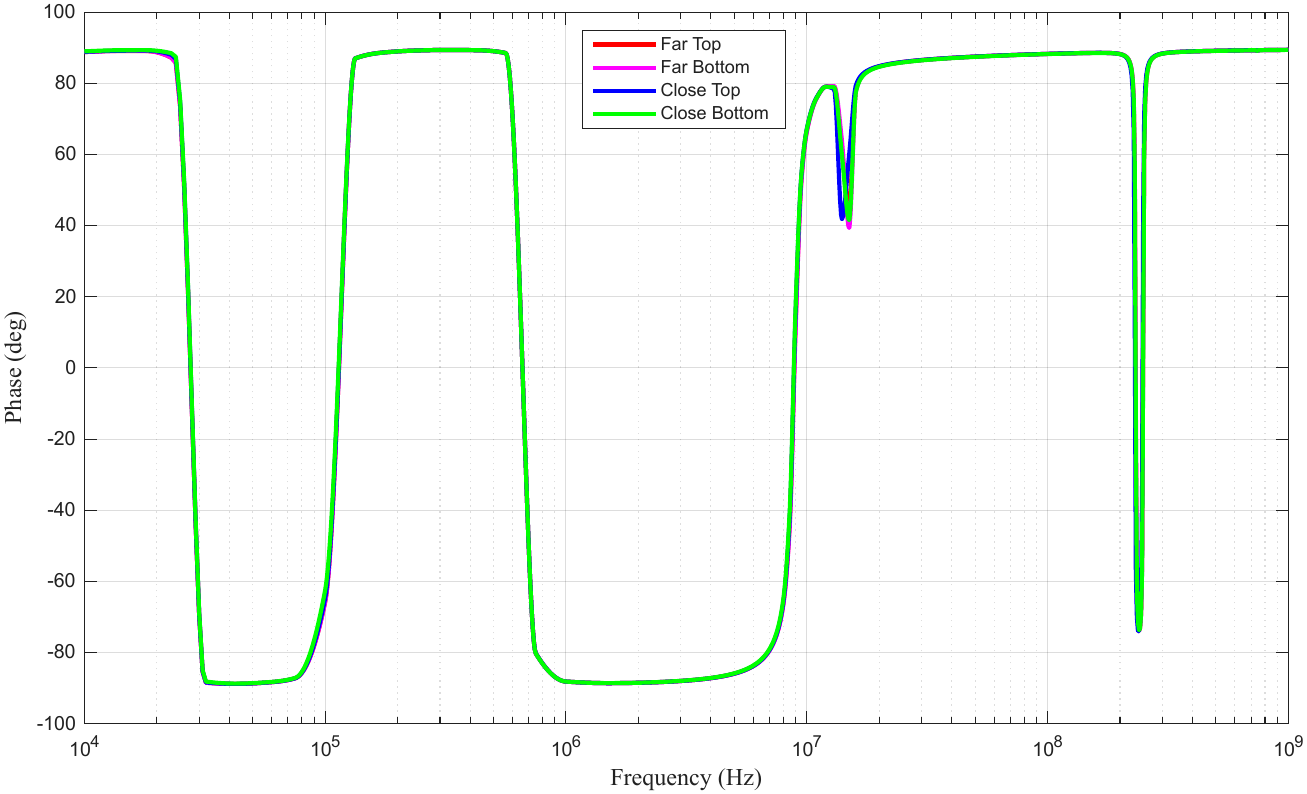}{d}
\caption{Effect of the position of the VRMs. (a) The four simulated positions. (b) Impedance comparison. (c) Imaginary impedance comparison. (d) Phase comparison.\label{fig:22}}
\end{figure*}

In this part, the comparison is widened beyond lateral and vertical placement, so that the impedance can be compared when the VRMs are placed in different locations on the motherboard. All four positions are shown in Fig.~\ref{fig:22}(a), namely laterally far on the top side, laterally close to the SoC on the top side, laterally far on the bottom side, and vertically below the SoC. The mid- and high-frequency decoupling capacitors are left unchanged.

Fig.~\ref{fig:22}(b) shows again that only the low-frequency impedance changes, since the loop inductance of the other decoupling capacitors is not affected. The lateral positions at the top and at the bottom of the board produce almost the same loop inductance. These distant lateral placements give the worst impedance of the set; the lateral VRMs closer to the die give a better response over the low-frequency range; and the vertical placement gives the lowest impedance of all. The imaginary impedance in Fig.~\ref{fig:22}(c) leads to the same ranking, since the inductive impedance is worst in the two distant lateral cases, better in the closer lateral case, and best in the vertical case.

The phase is examined in Fig.~\ref{fig:22}(d), because the component values are identical in all four simulations and only the length of the loop differs. Almost no discrepancy is found between the four positions. The comparison with Fig.~\ref{fig:22}(c) shows that the phase is about $+90^{\circ}$ wherever the board presents an inductive impedance, as the behavior of an inductor requires, and about $-90^{\circ}$ in the capacitive regions. Fig.~\ref{fig:22}(d) also reveals the inductive and the capacitive, or diffusive, response of the power and ground planes, which does not appear in the impedance or in the imaginary impedance plots, because this impedance is so low that it cannot be distinguished there when the impedance is not plotted on a logarithmic scale.

\subsection{Power and Ground Planes Position in LPD}
\begin{figure*}[!t]
\centering
\begin{minipage}[c]{0.49\textwidth}\centering
\panel{\linewidth}{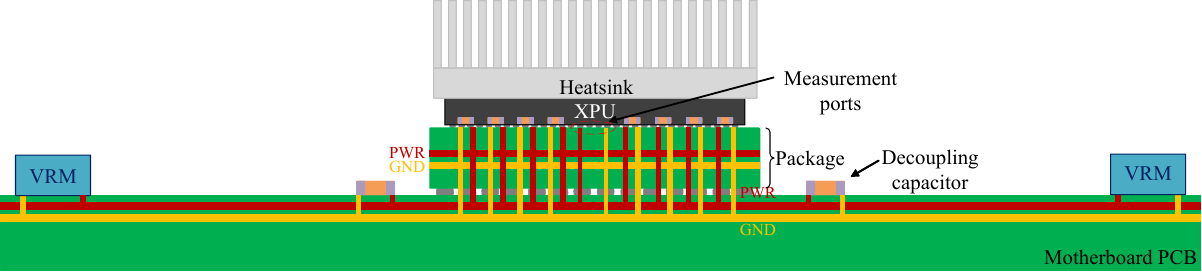}{a}\\[2pt]
\panel{\linewidth}{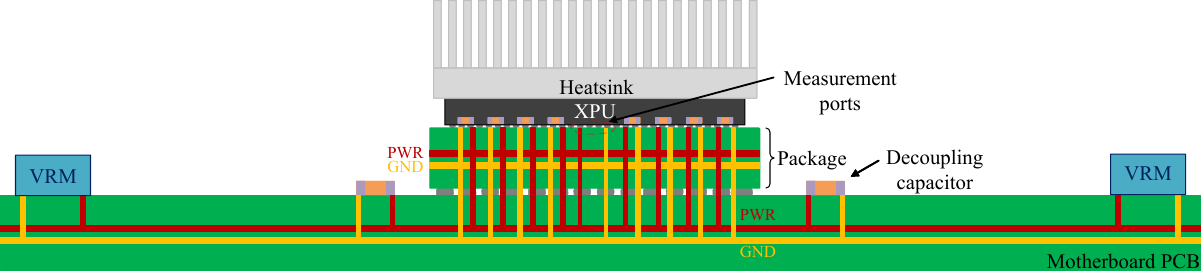}{b}\\[2pt]
\panel{\linewidth}{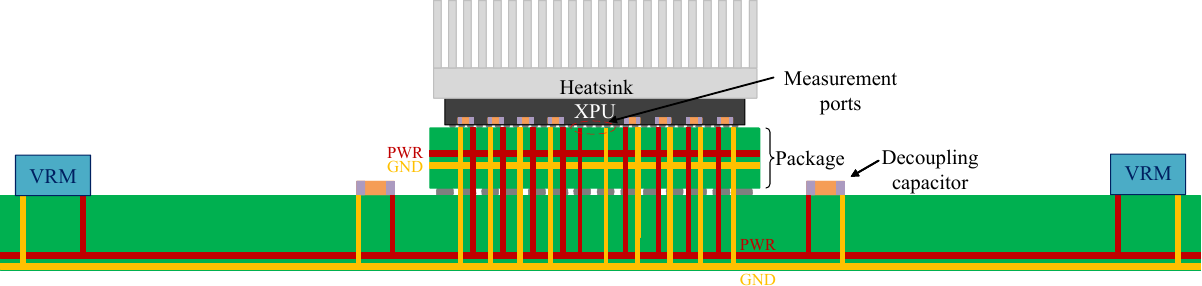}{c}
\end{minipage}\hfill
\panel{0.49\textwidth}{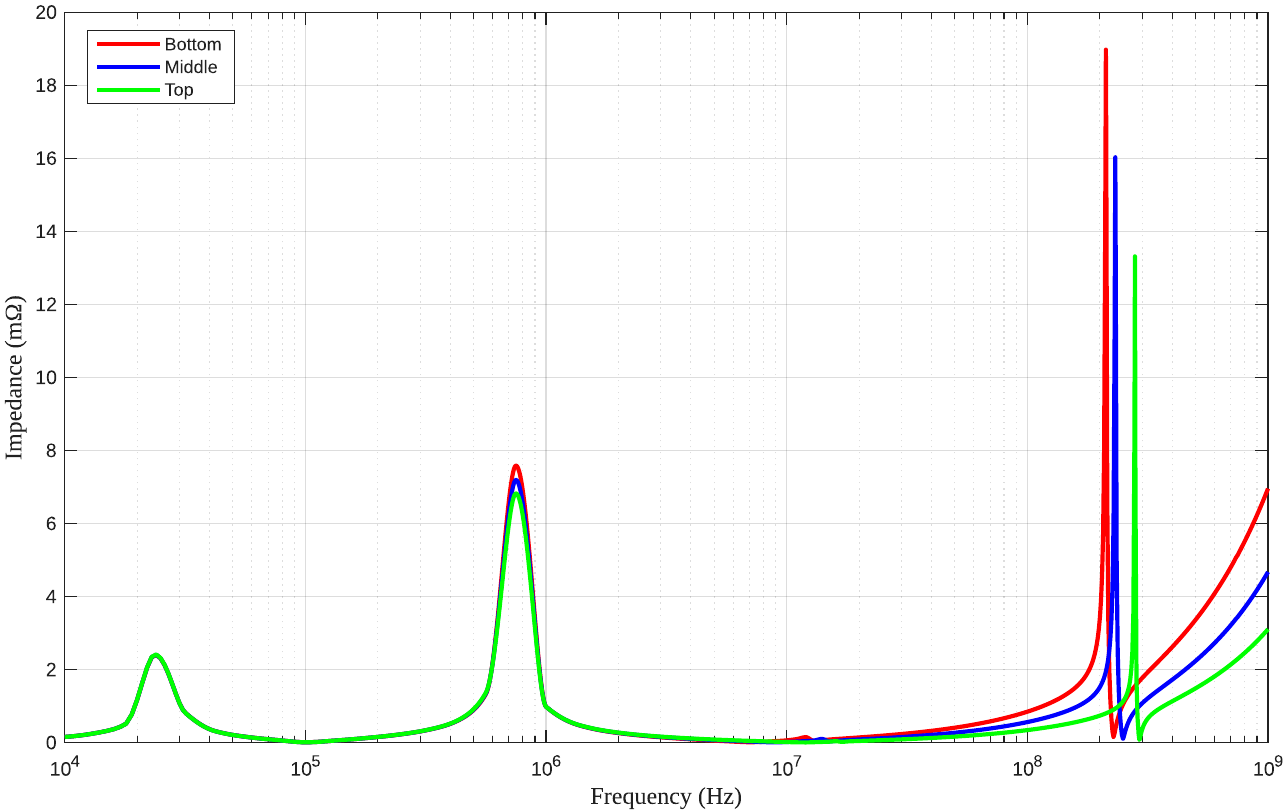}{d}\\[4pt]
\panel{0.49\textwidth}{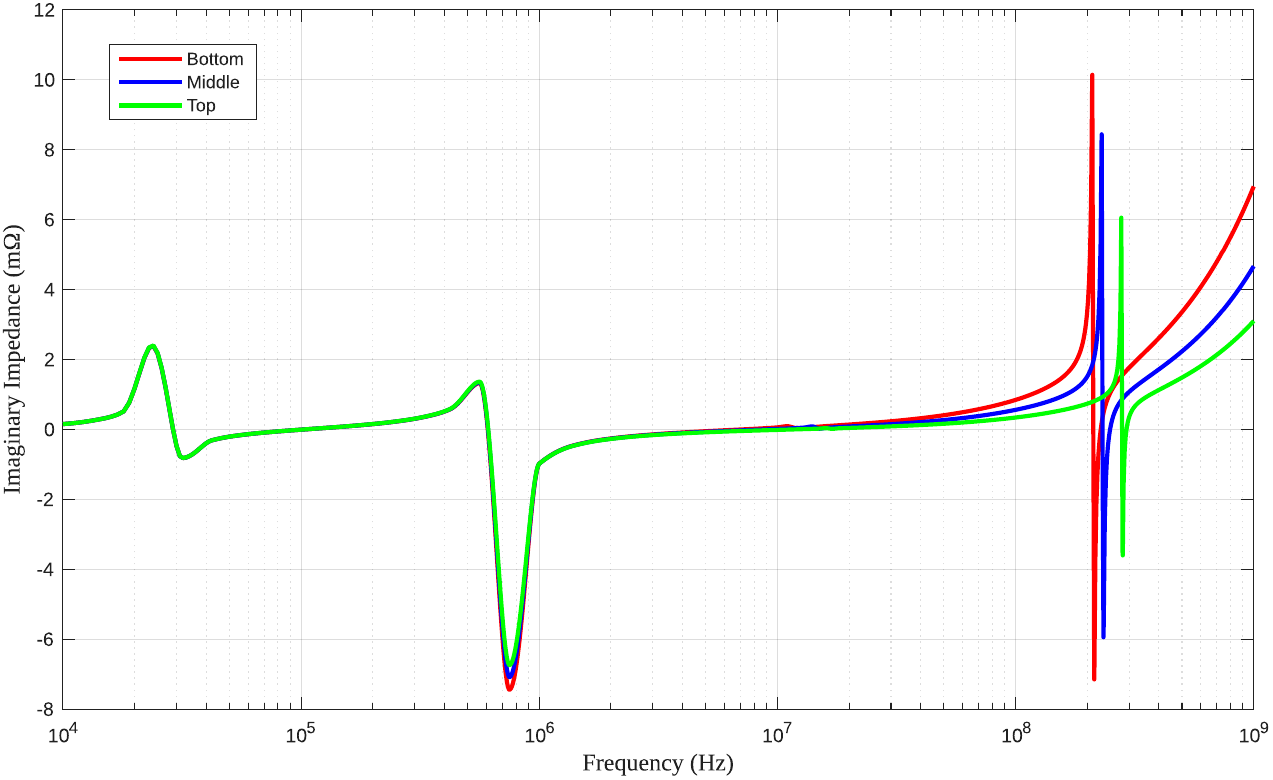}{e}\hfill\panel{0.49\textwidth}{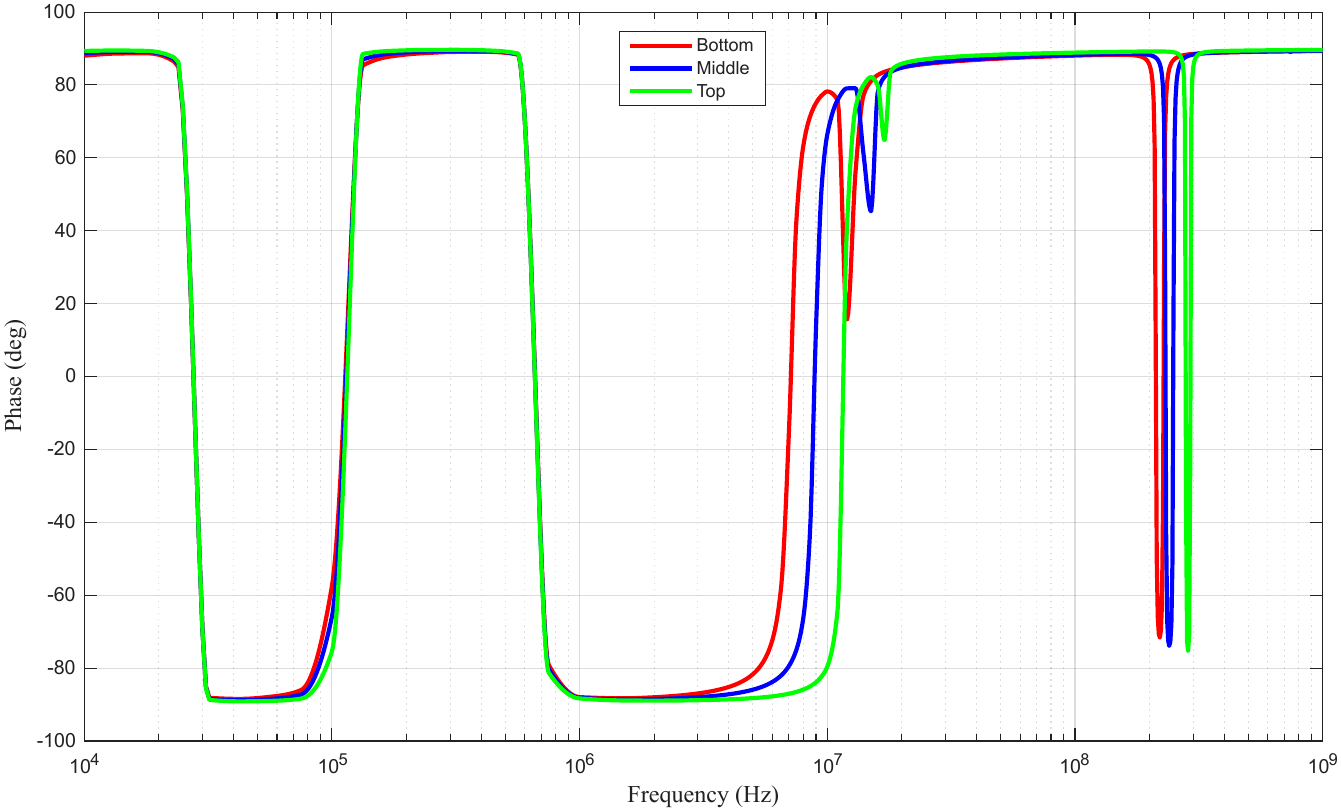}{f}
\caption{Effect of the position of the power and ground planes in LPD. (a) Planes at the top. (b) Planes at the middle. (c) Planes at the bottom. (d) Impedance comparison. (e) Imaginary impedance comparison. (f) Phase comparison.\label{fig:23}}
\end{figure*}

The impedance seen by the PDN is changed not only by the lateral or vertical placement of a VRM, but also by the relative position of the power and ground planes to which the vias of that VRM connect. Three plane positions are therefore simulated for each delivery scheme, with the same spacing used in the lateral and in the vertical case. The lateral layouts are shown in Fig.~\ref{fig:23}(a) for the top planes, which lie closest to the die, in Fig.~\ref{fig:23}(b) for the planes at the middle of the board, and in Fig.~\ref{fig:23}(c) for the planes at the bottom.

The three impedances are compared in Fig.~\ref{fig:23}(d). No change appears in the low-frequency section, whereas the mid- and the high-frequency ranges both change, and the reason lies in the size of each loop relative to the distance between the capacitors and the measurement ports at the die. In the low-frequency range the loop inductance grows only slightly as the planes move away from the VRM and the die. For the mid-frequency decoupling capacitors, the same increase in loop distance weighs more heavily, since these capacitors already stand close to the die, and Fig.~\ref{fig:23}(d) shows this effect beginning to appear, as the impedance over the mid-frequency range rises with the distance between the planes and the VRM and die. The largest difference is found in the high-frequency range, where the loop distance becomes much greater as the planes move away. Here the change in loop size shifts both the high-frequency anti-resonant and resonant frequencies, and the closer the planes lie to the decoupling capacitors and the die, the lower the impedance becomes, again because the loop inductance is lower.

The imaginary impedance of the three plane positions, compared in Fig.~\ref{fig:23}(e), shows the same effect as the magnitude. The phase in Fig.~\ref{fig:23}(f) separates two distinct frequency ranges in which different reactive effects take place. The first belongs to the power and ground planes themselves and appears between about 2 and 20~MHz, where the diffusive response of the planes becomes more noticeable as their distance from the measurement port increases. The second appears in the high-frequency range and is caused by the higher loop inductance of the more distant planes, which makes the anti-resonant and resonant frequencies differ.

\subsection{Power and Ground Planes Position in VPD}
\begin{figure*}[!t]
\centering
\begin{minipage}[c]{0.49\textwidth}\centering
\panel{\linewidth}{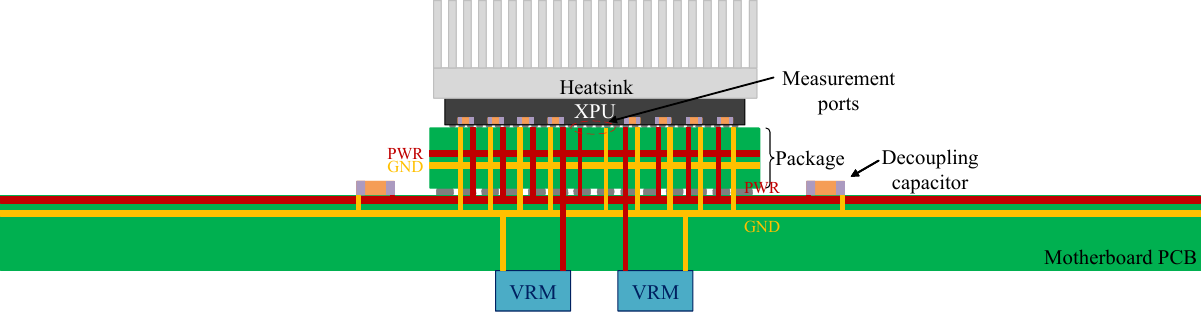}{a}\\[2pt]
\panel{\linewidth}{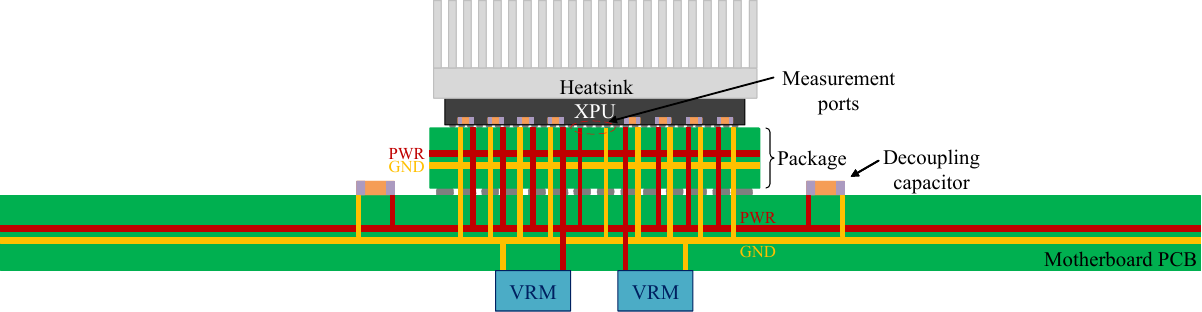}{b}\\[2pt]
\panel{\linewidth}{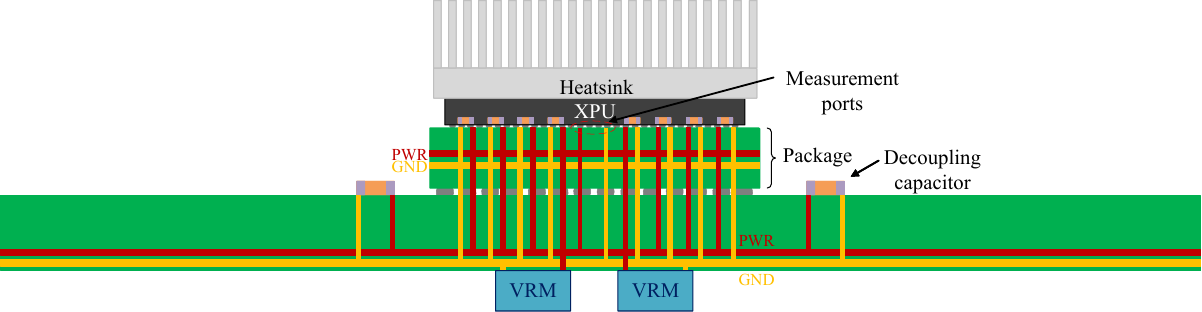}{c}
\end{minipage}\hfill
\panel{0.49\textwidth}{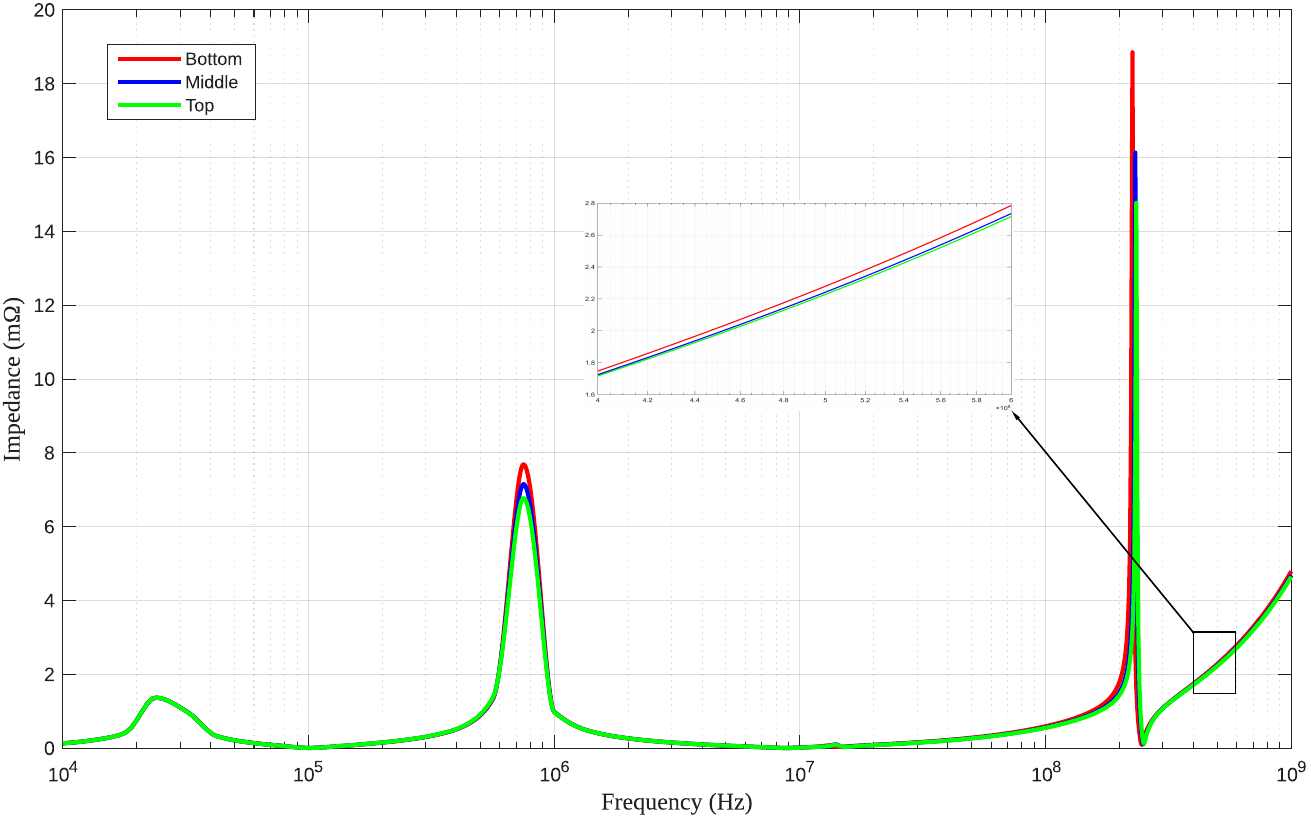}{d}\\[4pt]
\panel{0.49\textwidth}{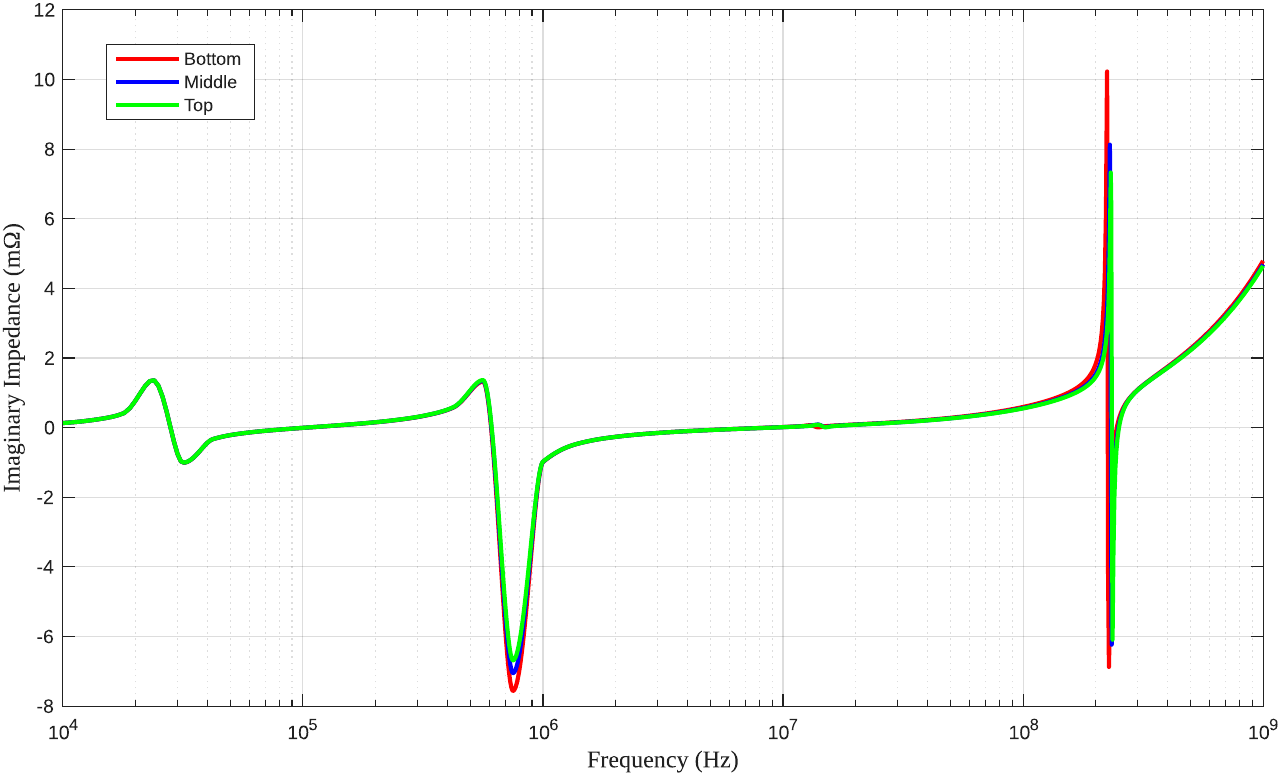}{e}\hfill\panel{0.49\textwidth}{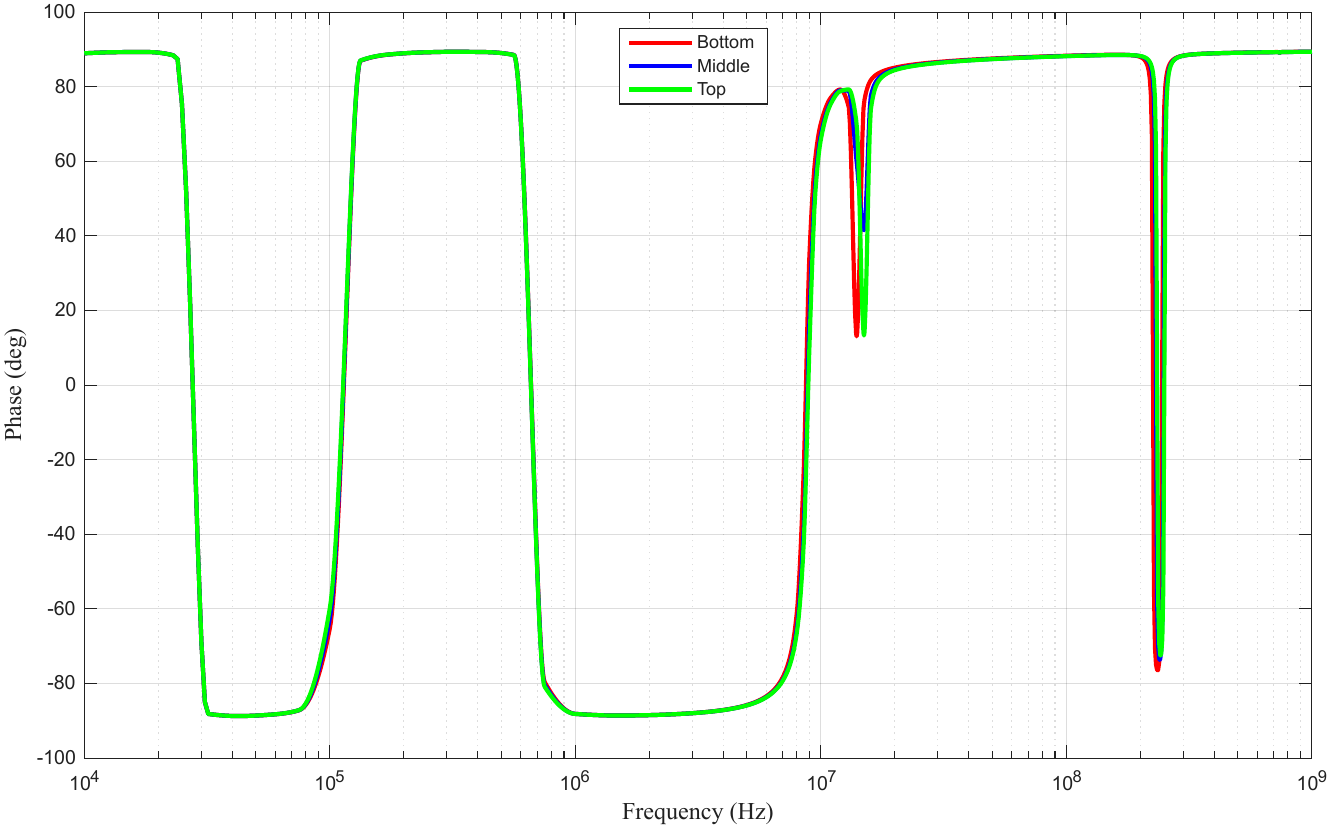}{f}
\caption{Effect of the position of the power and ground planes in VPD. (a) Planes at the top. (b) Planes at the middle. (c) Planes at the bottom. (d) Impedance comparison. (e) Imaginary impedance comparison. (f) Phase comparison.\label{fig:24}}
\end{figure*}

The other set of simulations repeats the same study for VPD, so that the impedance can be compared as the planes are brought closer to or moved away from the measurement ports at the die. Fig.~\ref{fig:24}(a) shows the planes at the top position closest to the die, Fig.~\ref{fig:24}(b) at the middle of the PCB, and Fig.~\ref{fig:24}(c) at the bottom, which is the position farthest from the die.

The first difference between Fig.~\ref{fig:23}(d) and Fig.~\ref{fig:24}(d) is that the low-frequency impedance is lower in the vertical board. A similar change to the lateral case is found in the mid-frequency range, where the impedance becomes higher as the planes move away from the die. The high-frequency anti-resonant peaks and resonant troughs also differ between the two boards, and those of Fig.~\ref{fig:24}(d) lie closer together than those of Fig.~\ref{fig:23}(d), because the loop of the low-frequency vertical VRMs does not change with the plane position and is smaller than in the lateral case. The imaginary impedance in Fig.~\ref{fig:24}(e) shows the same picture, since the low-frequency imaginary impedance is lower for VPD than for LPD, and the mid- and high-frequency ranges differ for the same reason as in Fig.~\ref{fig:24}(d). Finally, the phases of the three plane positions are compared in Fig.~\ref{fig:24}(f), where both the diffusive response of the planes and the phase changes at the high-frequency anti-resonance lie closer together than in the lateral case of Fig.~\ref{fig:23}(f). This occurs for the same reason as the impedances, namely that the high-frequency anti-resonant impedance of the vertical board is less affected by changes in the loop of the VRMs.

\subsection{Design Guidelines for Better PDN Behavior}
The behavior observed in these simulations can be expressed as a set of rules for the design of a PDN.
\begin{enumerate}
\item \textit{Place each stage according to the band it serves}: Moving the regulator improves only the part of the spectrum that this regulator supplies. Relocating the VRMs below the die lowered the low-frequency anti-resonance and left the mid- and high-frequency ranges untouched, so a lower impedance at mid and high frequency has to be sought from the stages that serve those ranges.
\item \textit{Treat the loop, not the component, as the design variable}: The component values were identical in every comparison, and all of the improvement came from the size of the current loop. The loop inductance is therefore the quantity to compare when two placements are judged.
\item \textit{Bring the regulator under the die wherever it is possible}: The vertical position gave the lowest impedance and the lowest inductive reactance of all the positions examined, and it lowered the anti-resonant peak, which is the feature that pushes the impedance profile through a target mask.
\item \textit{Where vertical placement is not possible, shorten the lateral path}: The lateral VRMs at the top and at the bottom of the board gave almost identical responses, while the lateral VRMs closer to the die gave a clearly better one. What matters is the length of the loop and not the side of the board on which the regulator is placed.
\item \textit{Keep the power and ground planes close to the die and to the regulator}: The plane position left the low-frequency range unchanged but raised the impedance over the mid-frequency range and shifted the high-frequency anti-resonant and resonant frequencies. The planes should therefore be kept as close to the die as possible.
\item \textit{Expect vertical delivery to make the high-frequency behavior more stable}: In the vertical board the high-frequency peaks and troughs moved less when the planes were moved, because the loop of the vertical VRMs is short and does not change. A vertical arrangement is thus less sensitive to the layer assignment than a lateral one.
\item \textit{Place decoupling capacitors as close to the die as possible}: Place the decoupling capacitors below or as close to the die as possible to reduce the parasitic loop inductance.
\item \textit{Add more vias to decrease path impedance}: Adding vias around the components decreases the overall impedance by allowing more routes and thus improves power and signal integrity.
\item \textit{Use thinner power/ground planes}: The thinner the power/ground planes are, the better the frequency performance will be.
\end{enumerate}

\section{Conclusion}
\label{sec:conc}
This paper reviews the PDN of AI data centers and the transition from LPD to VPD. The reviewed literature indicates that this transition is being driven by rising processor current, fast load transients, and the growing limitations of LPD in voltage drop, copper loss, board area, and power pin count. These limitations grow with the load current and cannot be removed by improving the converters alone. The 800~VDC architecture was also reviewed, since it brings the conversion stages closer to the GPU and allows the two-stage VRM to be replaced by a single-stage VRM, which can be implemented with either LPD or VPD. In this context, VPD has been presented as a promising approach for shortening the delivery path, lowering its resistance and inductance, and freeing the board area around the processor.

The PDN was reviewed as a hierarchy of charge sources whose impedance must stay below a target mask, and the position of the regulator was shown to shape the impedance profile of the whole network. VPD implementation approaches, reported converter designs, and packaging technologies were then examined. In separate designs, the reviewed converters reach heights as low as 2.6~mm and current densities as high as 0.71~A/mm$^{2}$. IVRs, FIVRs, and embedded passive components were also reviewed, since they shorten the delivery path further. An on-die FIVR leaves the PDN with a single anti-resonant frequency, so that the rogue wave effect can no longer occur. The cost is an inductor that must be designed for the package, capacitors that must be built into the silicon, and heat that is difficult to remove from the regulator.

Power integrity simulations were carried out in Cadence Sigrity PowerSI to examine these claims. In a complete PDN, relocating the VRMs below the die lowered the low-frequency anti-resonant peak and left the mid- and high-frequency ranges unchanged. The lateral positions at the top and at the bottom of the board gave almost identical responses, which shows that the length of the loop matters rather than the side of the board. Moving the power and ground planes away from the die left the low-frequency range unchanged, but it raised the mid-frequency impedance and shifted the high-frequency anti-resonant and resonant frequencies. These results led to a set of PDN design guidelines: each stage should be placed according to the band it serves, the loop should be treated as the design variable, the regulator should be brought under the die wherever possible, the lateral path should be shortened where it is not, and the planes should be kept close to the die and the regulator. Since each VRM was represented by its output capacitor, these results describe the passive impedance of the network rather than the active response of the regulator. Overall, the literature suggests that future progress will depend on power delivery solutions that can bring the regulator closer to the die while addressing regulator height, passive-component embedding, and thermal management.

\end{document}